\documentclass[a4paper,fleqn,usenatbib]{mnras}

\usepackage[T1]{fontenc}
\usepackage{ae,aecompl}

\usepackage[pdftex]{graphicx}	% Including figure files
\usepackage{amsmath}	% Advanced maths commands
\usepackage{amssymb}	% Extra maths symbols

\newlength\figureheight 
\newlength\figurewidth 
\newcommand{\adjustThreePanels}{
    \setlength\figureheight{6cm}\ignorespaces 
    \setlength\figurewidth{6cm}\ignorespaces
    \setlength{\tabcolsep}{1pt}\ignorespaces
}

\usepackage{ctable}

\newcommand{\degrees}{\ensuremath{^\circ}}
\newcommand{\Msun}{\ensuremath{\mathrm{M_{\odot}}}}
\newcommand{\Rsun}{\ensuremath{\mathrm{R_{\odot}}}}

\newcommand{\Proteus}{{\sc Proteus}}
\newcommand{\FLASH}{{\sc FLASH}}

\title[The ejecta-companion interaction]
      {Imprints of the ejecta-companion interaction in Type Ia supernovae: main sequence, subgiant, and red giant companions}
\author[P. Boehner, T. Plewa and N. Langer]
  {P. Boehner$^1$\thanks{E-mail: philboehner@gmail.com (PB);
      tplewa@fsu.edu (TP); nlanger@astro.uni-bonn.de (NL)},
   T. Plewa$^{1\ast}$, 
   N. Langer$^{2\ast}$\\
  $^1$Department of Scientific Computing, Florida State University, Tallahassee, FL 32306, U.S.A.,\\
  $^2$Argelander-Institut f\"ur Astronomie der Universit\"at Bonn, Aufdem H\"ugel, D-53121 Bonn, Germany}
\begin{document}

\date{Accepted 2016 October 20. Received 2016 October 11; in original form 2016 August 17}

\pagerange{{2060}--{2075}} \pubyear{2017} \volume{465}

\maketitle

\label{firstpage}
\begin{abstract}
We study supernova ejecta-companion interactions in a sample of
realistic semidetached binary systems representative of Type Ia
supernova progenitor binaries in a single-degenerate scenario. We
model the interaction process with the help of a high-resolution
hydrodynamic code assuming cylindrical symmetry. We find that the
ejecta hole has a half-opening angle of 40--50\degrees\ with the density by a
factor of 2--4 lower, in good agreement with the previous studies.
Quantitative differences from the past results in the amounts and
kinematics of the stripped companion material and levels of
contamination of the companion with the ejecta material can be
explained by different model assumptions and effects due to numerical
diffusion.We analyse and, for the first time, provide simulation-based
estimates of the amounts and of the thermal characteristics of the
shock-heated material responsible for producing a prompt, soft X-ray
emission. Besides the shocked ejecta material, considered in the
original model by Kasen, we also account for the stripped,
shock-heated envelope material of stellar companions, which we predict
partially contributes to the prompt emission. The amount of the energy
deposited in the envelope is comparable to the energy stored in the
ejecta. The total energy budget available for the prompt emission is
by a factor of about 2--4 smaller than originally predicted by
Kasen. Although the shocked envelope has a higher characteristic
temperature than the shocked ejecta, the temperature estimates of the
shocked material are in good agreement with the Kasen's model. The
hottest shocked plasma is produced in the subgiant companion case.

\end{abstract}
\section{Introduction}
Type Ia supernovae (SNeIa), are used as 'standard candles' to measure
cosmological distances \citep{Phillips+93} and the expansion rate of
the universe \citep{Branch+92}. However, the fundamental processes
leading up to these events are still not fully understood \citep[see,
  e.g.][and references therein]{Hillebrandt+00}. Though much work has
gone into improving the procedure used to standardize the SNIa light
curves, this method is mainly empirical and does not take into account
the physical properties of the exploding objects. It is conceivable
that individual supernova events will be characterized with different
energies and nucleosynthetic yields depending on the actual state of
the exploding object. In the Type Ia single degenerate (SD) scenario
\citep[]{Whelan+73,Nomoto+82a}, the explosion involves a massive white
dwarf while in the double degenerate (DD) scenario
\citep[]{Iben+84,Webbink+84}, two white dwarfs in a close binary
system violently merge in a process that is expected to eventually
produce an explosion.

Identifying which specific scenario produced a particular Type Ia
supernova requires precise observations. For example, no companion
star should be found near the center of the explosion site following
the SNIa explosion in the binary white dwarf system. In contrast, one
expects to observe a late-type star in the immediate vicinity of the
Type Ia supernova produced in the SD channel. The observations of
field stars in the central region of Tycho supernova remnant have
provided the strongest evidence for such a late-type companion star
\citep[]{Ruiz-Lapuente+04,Ruiz-Lapuente+14}, although the reported
discovery remains controversial \citep[]{Xue+15,Williams+16}. In
addition, the imaging and spectroscopic observations of the remnant
may provide evidence for interaction between the progenitor system
with the interstellar medium \citep[]{Zhou+16}. Some theoretical
models predict supernova ejecta to show morphological or compositional
features that could be linked to the presence of a non-degenerate
companion star \citep[see, e.g.][]{Marietta+00}.

The process of the interaction between the supernova and a late-type
stellar companion should also produce a substantial amount of shocked
material, which could produce excess radiation in the UV and possibly
also in the soft X-rays. Detection of such excess emission was
recently reported in the Type Ia supernovae SN iPTF14atg
\citep{Cao+15} and SN 2012cg \citep{Marion+16}. The interpretation of
the observed excess emission follows the theoretical model by
\citet{Kasen+10}, who was the first to consider the emission
produced in the process of collision between the Type Ia supernova
ejecta with various non-degenerate stellar companions.

Such interactions were originally modeled by means of hydrodynamic
simulations by \citet{Marietta+00}. These authors performed the
comprehensive study in two spatial dimensions using a high resolution
finite volume hydrocode. They considered a variety of plausible
single-degenerate progenitor systems with main sequence (MS), subgiant
(SG), and red giant (RG) companions. They quantified the amount of
hydrogen stripped from the companion, the velocity distribution of the
stripped mass, the momentum transferred by the ejecta to the companion
(the kick velocity), and the contamination of the companion with
supernova ejecta material. Their most notable prediction was the
existence of a low-density region carved in the ejecta by the
companion. Since then, this specific morphological feature of a
single-degenerate Type Ia supernova remnant is commonly known as the
``ejecta hole''.

The results of \citet{Marietta+00} were later scrutinized in three
dimensions by \citet{Pakmor+08}. This group used a smoothed particle
hydrodynamics (SPH) method and limited their studies to MS
companions. They considered the dependence of model outcomes on the
supernova explosion energy and the distance to the companion, and
found evidence for a power-law relation between the amount of stripped
mass and the kick velocity on the explosion energy based on the
analysis of one of their model binary systems. Their findings in
regard to the geometry of the ejecta hole largely confirmed the
original predictions of \citet{Marietta+00}.

More recently, \citet{Liu+12} further developed a SPH model of
\cite{Pakmor+08} and used improved MS companion models obtained in the
course of pre-supernova binary evolution \cite{Ivanova+04}. They found
much higher amounts of material stripped from companions than reported
by \cite{Pakmor+08}. Additionally, they found that although the amount
of stripped mass and kick velocity still depend on the impact geometry
(the ratio between the stellar companion radius and the orbital
separation), the actual relations also depend on the companion's
structure. These types of power law correlations, including the amount
of contamination as a function of the explosion energy, were later
reported by \citet{Liu+13b} also in the case of a helium companion.

\citet{Pan+10} were the first to employ an adaptive mesh refinement
technique to study the interaction process, and considered a binary
system with a helium companion in two dimensions. They found power-law
relations for the amount of stripped mass and kick velocity of the
companion similar to those originally reported by
\citet{Pakmor+08}. Their model was later extended in \citet{Pan+12a},
who accounted for the effects of orbital motion and the companion's
spin. In that more recent work, they performed an extensive parameter
survey in two and three dimensions varying binary orbital separations
for RG, MS, and helium companions. In regard to contamination of
companion stars with the ejecta material, they found that the amount
of contamination of their model MS companion was comparable to that
reported by \cite{Marietta+00}. However, in the case of a compact
helium companion, the level of contamination was an order of magnitude
greater than obtained by \citet{Marietta+00}. Additionally, they found
no substantial amount of supernova debris deposited onto their RG
model companion.

The evolution of the companion star after the interaction with the
supernova ejecta is addressed in more recent studies. For the
short-term evolution of the surviving companion star, \cite{Liu+13a}
studied MS companion stars in three dimensions for about 3 h, finding
that the rotational velocity is reduced to about a fourth of the
pre-supernova velocity due to expansion and the fact that much of the
angular momentum was carried away by the stripped mass. For the
long-term evolution, \cite{Pan+12b,Pan+13}, using the methods and
results of \cite{Pan+12a} as input for a one-dimensional stellar
evolution code for MS, SG, and helium stars, found that the
rotational velocity of the companion star will be reduced and the star
will enter a highly luminous phase years following the supernova
event. \cite{Pan+14} then found an upper limit in terms of detection
for the distance a MS or helium star could travel after
being kicked from the explosion site.

The long-term evolution of the ejecta hole was studied by
\citet{Garcia-Senz+12}, using an SPH technique. They considered a
single system with a main-sequence companion and accounted for the
effects of orbital motion. They conjectured that the hole will likely
close on the time scale of centuries due to hydrodynamic
instabilities. It should be noted, however, that these authors did not
consider the role of perturbations existing in the interstellar
medium, such as turbulence, that would likely contribute to the
destruction of the ejecta hole and mixing of the material stripped
from the companion with the ejecta.

Our aim is to extend the above studies by performing hydrodynamic
simulations of the supernova ejecta-companion interaction for a rich
set of binary systems considered to be realistic progenitor systems of
Type Ia supernovae in the single degenerate channel. We discuss a
range of basic model outputs such as the structure of the ejecta hole,
kinematic properties of the stripped companion material, and pollution
of the supernova companion stars by the supernova ejecta, and compare
them to the results obtained by other groups. More importantly, and in
the context of the prompt soft X-ray emission model, we carefully
analyze and, for the first time, provide information about the amounts
and thermal characteristics of the energy stored in the ejecta {\emph and}
in the envelope of the companion by shocks created due to
collision between the ejecta on the companion's envelope. The
contribution of the shocked envelope material to the prompt emission
was not included in the original model by \citet{Kasen+10}. Our
results provide input to more realistic radiative transport
calculations of the prompt emission that should help discerning
different types of companion stars using the multi-wavelength
photometric observations.
\section{Models and Methods} \label{s:methods-models}
We begin our presentation with discussion of models of companion
stars, parameters of binary systems, and the supernova explosion model
used in our study. These elements determine our choice of numerical
methods and parameters of computer models, including hydrodynamic
solvers and numerical discretization.
\subsection{Stellar physics input} \label{ss:stellar-input}
\subsubsection{Stellar companions and corresponding binary system models} \label{ss:models}
Due to the unknown nature of supernova companions, it is important to
consider various feasible binary systems in order to obtain a broad
spectrum of outcomes to enable comparison with observations. In this
work, we consider seven companion types: four MS stars, one SG star,
and two RG stars. Table \ref{t:simParams}
%
% Add the radial extent of each model before t_i
%
\ctable[
    cap = model binary systems table,
    caption = { Parameters of model binary systems and of their computational models.},
    label = t:simParams,
    star,
    ]
    {l c c c c c c c c c}
    {
       \tnote[a]{Mass of the binary companion.}
       \tnote[b]{Mass of the point mass used to represent the degenerate core of the RG companion.}
       \tnote[c]{Radius of the binary companion.}
       \tnote[d]{Binary separation measured from the center of the supernova to the center of the companion.}
       \tnote[e]{Orbital period of the progenitor system.}
       \tnote[f]{Impact parameter.}
       \tnote[g]{Radial extent of the domain.}
       \tnote[h]{Initial simulated time.}
       \tnote[i]{Final simulated time.}
    }
    {\FL
    Model       & $m_{\ast}$\tmark[a]    & $m_\mathrm{pm}$\tmark[b] & $R_{\ast}$\tmark[c]    & $a$\tmark[d]          & $P$\tmark[e]  & $R_{\ast}$/$a$\tmark[f] & $r_{max}$\tmark[g]  & $t_i$\tmark[h] & $t_f$\tmark[i]              \NN
    designation & [M$_{\odot}$] & [M$_{\odot}$]         &    [R$_{\odot}$] & [R$_{\odot}$] & [d]   &               & [cm] & [s]     & [s]              \ML
    MS7   & 1.53                     & --    & 2.57                      & 6.58          & 1.15   & 0.39         & 5.86 $\times$ 10$^{13}$ & 8.80 $\times$ 10$^{1}$ & 2.00 $\times$ 10$^{4}$     \NN 
    MS38  & 1.15                     & --    & 1.07                      & 2.94          & 0.37   & 0.37         & 4.89 $\times$ 10$^{13}$ & 4.00 $\times$ 10$^{1}$ & 1.98 $\times$ 10$^{6}$     \NN
    MS54  & 1.24                     & --    & 0.79                      & 2.13          & 0.22   & 0.37         & 3.61 $\times$ 10$^{13}$ & 2.80 $\times$ 10$^{1}$ & 2.00 $\times$ 10$^{4}$     \NN
    MS63  & 1.13                     & --    & 1.40                      & 3.84          & 0.55   & 0.36         & 6.37 $\times$ 10$^{13}$ & 5.30 $\times$ 10$^{1}$ & 2.00 $\times$ 10$^{4}$     \NN
    SG    & 1.53                     & --    & 3.18                      & 8.15          & 1.59   & 0.39         & 7.25 $\times$ 10$^{13}$ & 1.10 $\times$ 10$^{2}$ & 4.32 $\times$ 10$^{4}$     \NN
    RG319 & 0.61                     & 0.315 & 54.6                      & 175           & 192    & 0.31         & 2.49 $\times$ 10$^{15}$ & 2.80 $\times$ 10$^{3}$ & 8.64 $\times$ 10$^{5}$     \NN 
    RG428 & 0.75                     & 0.423 & 249                       & 755           & 1659   & 0.34         & 1.14 $\times$ 10$^{16}$ & 1.13 $\times$ 10$^{4}$ & 5.78 $\times$ 10$^{6}$     \LL
    }
presents the properties of the companions and their respective binary
systems. Our SG and RG model systems were obtained using the method of
\cite{Langer+00}, while our MS models were adopted from the existing
model database (see Table 2 in \cite{Langer+00}). For each model, we
show the impact parameter, $R_{\ast}/a$, where $R_{\ast}$ is the
radius of the companion and $a$ is the separation distance between the
centers of the supernova and the companion. The impact parameter is
similar for all systems considered, implying the orbital separation of
$a\approx 3 R$, similar to previous studies \citep[see,
  e.g.,][]{Marietta+00,Pakmor+08,Pan+12a}.

When selecting binary systems for our study, we aimed to cover a wide
range of masses and orbital parameters for our progenitor systems. For
example, the masses of our MS stars differ by about 50\%,
while their orbital separations (assuming the secondary fills its
Roche lobe) differ by a factor of more than 3. The SG star has a
mass equal to the most massive MS star, MS7, but is
somewhat larger due to advanced evolution and consequently has a wider
orbit. The two RG companions have significantly different radii and
orbital separations. It is worth noting that despite the differences
between the systems, the impact parameter differs only by about 30\%.
\subsubsection{Supernova explosion model}\label{ss:W7}
We use a W7-like supernova model \citep{Nomoto+84} to describe the
structure of a exploding Chandrasekhar-mass, carbon-oxygen white
dwarf. This model has been shown to closely match typical
characteristics of observed SNIa events in terms of energetics and
nucleosynthetic yields. Our model has an explosion energy of $\approx
1.15\times 10^{51}$ erg, with the velocity of the ejecta linearly
increasing with radius from its center to around 30000~km~s$^{-1}$ at 
its edge. The
ejecta is chemically stratified with the core region dominated by the
iron group elements, which is surrounded by a layer rich in
intermediate mass elements. The outermost ejecta layers are composed
of unburned carbon and oxygen.
\subsection{Computational methods} \label{ss:comp-methods}
It is interesting to note that despite continuous development of
supernova-companion interaction models, the main conclusions of the
original interaction model of \citet{Marietta+00} still
stand. Taking into account three-dimensional effects \citep[see,
  e.g.,][]{Pakmor+08,Garcia-Senz+12} or effects due to orbital motion
and companion's spin \citep{Pan+12a,Liu+12,Liu+13a} proved to have
only minor impact on the conclusions. Consequently, we decided to
model the supernova-companion interactions in two spatial
dimensions. Using this approach, we were able to study the evolution
of the binary systems on large domains for relatively long times and
at high resolutions. In what follows, we introduce our computational
model including the hydro solver and mesh discretization, and describe
the assumed initial and boundary conditions.

\subsection{Hydrodynamics and relevant physics} \label{ss:hydro}
In our study of ejecta-companion interactions, we use {\Proteus}, a
customized version of the \FLASH\ 2.4 code \citep{fryxell+00} for the
current application. \FLASH\ is a finite
volume multi-dimensional hydrodynamic block-structured AMR code. We
solve the Euler equations using the PPM solver \citep{colella+84}. In
addition to solving transport equations for mass, momentum, and total
energy, we also solve a set of equations describing advection of
nuclear species and passive mass scalars. The chemical composition of
our models is described using an extended alpha network with 19
isotopes; we do not include nuclear burning. The mass scalars are used
to track the evolution of ejecta material, stellar material, and the
ambient medium. We close the hydrodynamic system with the Helmholtz
equation of state \citep{timmes+00}. Finally, self-gravity is
calculated using the multipole expansion of the Poisson equation with
10 moments. 

In the study presented here, {\Proteus} differed from the original
\FLASH\ code in regards to mesh refinement, treatment of passively
advected mass tracers, and calculation of the gravitational
potential. The latter modification was required due to the limited
mesh resolution in our models, which required replacing the degenerate
cores of RG companions with point masses.  (The values of the point
masses, $m_\mathrm{pm}$, for RG319 and RG428 are given in Table
\ref{t:simParams} in solar masses.)  This procedure was not necessary
in the case of SG or MS companions.
\subsection{Computational domain and discretization} \label{ss:grid}
We simulate evolution of the binary systems in cylindrical geometry in
two dimensions assuming axisymmetry. Thus, our computational domain
starts at $r = 0$ and extends to a maximum radius $r_{max} = d$, with
the domain extending from $z_{min} = -d$ to $z_{max} = +d$ in the
z-direction. The characteristic size of the computational box, $d$, is
about 10$^{13}$ cm for MS and SG models, and between 10$^{15}$ and
10$^{16}$ cm for the RG models (see Table \ref{t:simParams} for
detailed information of the sizes of the computational domains). We
impose reflective boundary conditions along the symmetry axis, and
allow for free outflow elsewhere. We assume isolated boundary
conditions when calculating the gravitational potential.

We use an adaptive mesh discretization scheme wherein the mesh
resolution is increased whenever the local variation of density
exceeds $0.1$ or pressure exceeds $0.2$. The effective maximum mesh
resolution of our simulations is defined as the number of cells per
companion radius at the initial time. For every model family, we
performed a series of simulations with resolution gradually increasing
from 50 to 200 cells per companion radius. In what follows, we use a
notation in which, for example, a model with 200 cells per companion 
radius is denoted as L200.
\subsection{Initial conditions} \label{ss:ICs}
We constructed the initial conditions in our simulations by mapping
the supernova ejecta and the companion models onto the computational
domain filled with an ambient gas of density of $1 \times 10^{-16}$
g/cm$^{-3}$ and temperature of $1 \times 10^{4}$ K. Mapping of the
models to the mesh involved interpolation of density, temperature, and
chemical composition. As we discussed earlier in Section
\ref{ss:hydro}, both the internal energy and the pressure were
calculated using the Helmholtz equation of state.

The supernova was centered at the origin of our coordinate system with
a radius scaled in such a way that it nearly filled its Roche lobe. We
assumed that the supernova ejecta was isothermal with temperature of
$1 \times 10^{8}$ K. Also, as we discussed in Section \ref{ss:W7}, we
assumed that the ejecta was homologously expanding. The supernova
explosion energy was roughly equal to $1.15 \times 10^{51}$
erg. Additionally, we adjusted initial simulation times to account for
the geometrical expansion of the supernova model. In this way,
simulation times for various models can be directly compared with
$t=0$ corresponding to the supernova explosion time.

The stellar companion was placed in the positive direction along the
symmetry axis at the distance from the origin defined by the orbital
distance of the binary system model (cf.\ Table
\ref{t:simParams}). Before being interpolated onto the mesh, the
original companion model was iteratively relaxed to hydrostatic
equilibrium. This procedure was necessary in order to eliminate
unbalanced pressure gradients due to the differences between the
equation of state used to obtain the original companion model and the
Helmholtz equation of state. The relaxed companion structure, along
with the supernova ejecta model, was then mapped down to the mesh.

\begin{figure*}
\adjustThreePanels\ignorespaces
\begin{center}
    \begin{tabular}{ccc}
        %        % \tikzImage{figures/results/MergerMorphPanel1_0812}{}*[a]\ignorespaces
        \includegraphics{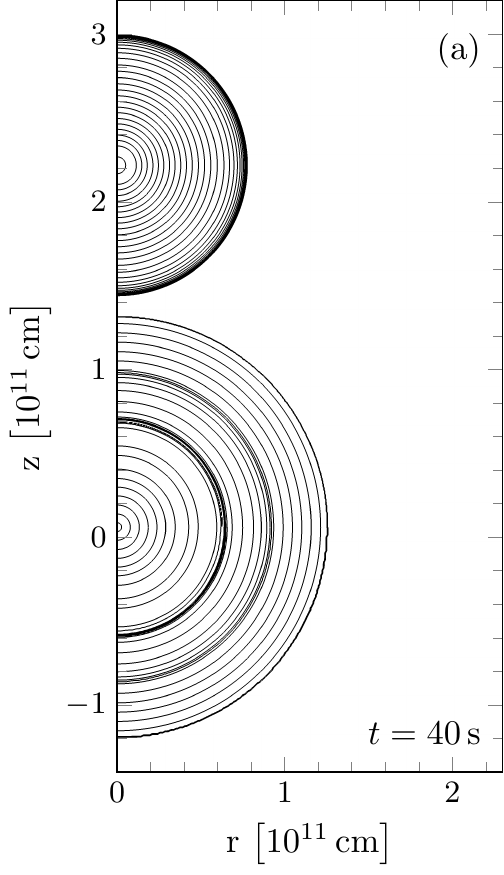}
        % \tikzImageContour{figs_publish/Fig1a_w7_ms38_logd_t_40s}{}*[a]\ignorespaces
        &
        \includegraphics{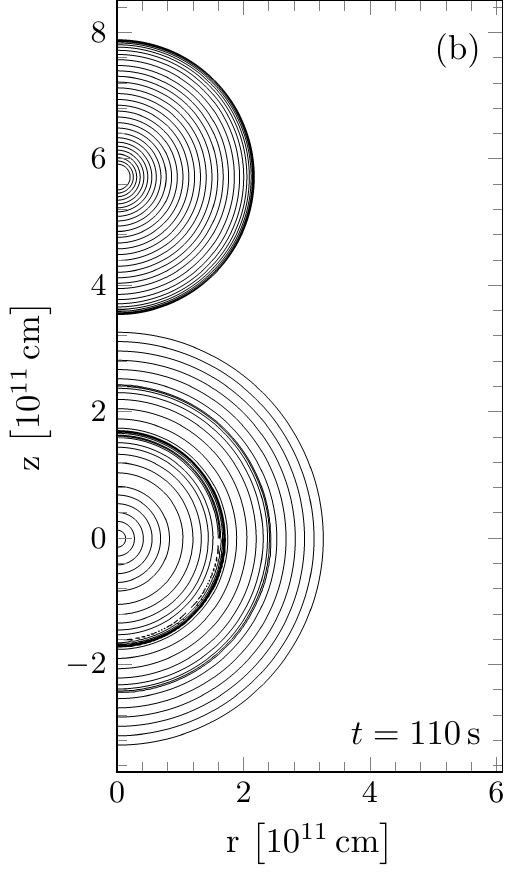}
        % \tikzImageContour{figs_publish/Fig1b_w7_sg_logd_t_110s}{}*[b]\ignorespaces
        &
        \includegraphics{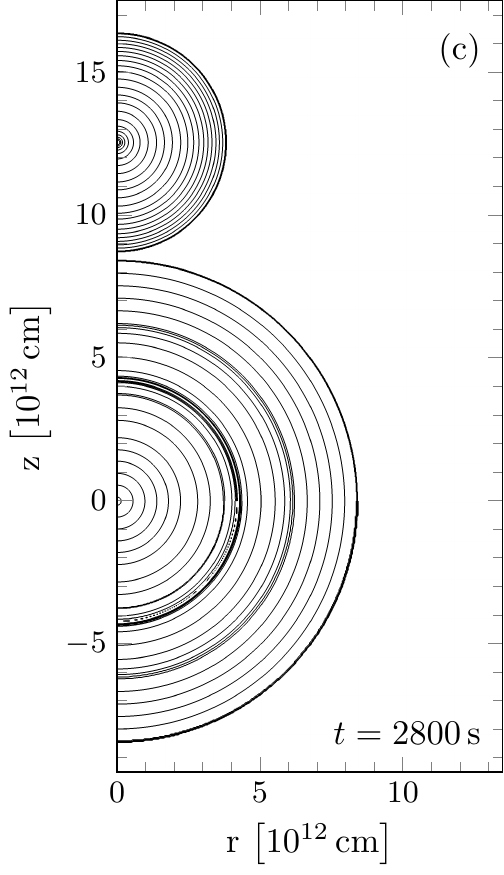}
        % \tikzImageContour{figs_publish/Fig1c_w7_rg319_logd_t_2800s}{}*[c]\ignorespaces
        %
    \end{tabular}
%
    % \blackLink{cb:mergerMorph0812Vert}
    % \includegraphics{figures/fig5colorbar.pdf}
% 
    \caption
    {Density contour maps of a selected subset of model binary systems
      at their initial simulation times. The density is shown with 30
      logarithmically spaced contour lines for (a) the main-sequence
      companion MS38 ($10^{-4}-10^{2}$ g cm$^{-3}$), (b) the subgiant
      SG ($10^{-5}-10^{1}$ g cm$^{-3}$), and (c) the red giant RG319
      ($10^{-10}-10^{-3}$ g cm$^{-3}$). Note that the initial
      simulation times and domain sizes differ between the
      panels.\label{fig:initial_contour} }
\end{center}
\end{figure*}

Fig.\ \ref{fig:initial_contour} 
shows the density distribution at the initial time for the MS38, SG,
and RG319 binary systems. The density is shown with 30 contours
logarithmically spaced between the density values specifically chosen
for every system. The dynamic range of the density plot shown changes
between 6 orders of magnitude for the MS38
(Fig.\ \ref{fig:initial_contour}(a)) and SG
(Fig.\ \ref{fig:initial_contour}(b)) models, and 7 orders of magnitude
for the RG319 model (Fig.\ \ref{fig:initial_contour}(c)). Due to the
aforementioned pre-expansion of the supernova ejecta, the initial
times differ from 40 seconds for the MS38 model to almost 1 h for the
RG319 model.

The binary interaction models were evolved as long as the supernova
shock was fully contained inside the domain. Thanks to this
constraint, we can fully account for different materials (as tracked
by corresponding mass scalars) in our analysis. Also, the interior
pressure of the expanding ejecta is unaffected by rarefaction waves
that may be produced as a result of the shock crossing the domain
boundaries. The final times for each model are shown in Table
\ref{t:simParams}.
\section{Results}\label{s:results}
We present the results of our simulations using a representative
subset of our binary models for the MS companion system MS38; subgiant
system, SG; and red giant system, RG319. The evolution in the
remaining cases of MS and RG companions does not qualitatively differ
from those included in the subset. The models presented in this
section were obtained at the resolution of L200.

In our presentation, we focus on the general development of the flow,
the propagation of the supernova shock through the companion interior
and the surrounding circumstellar medium, and the process of mixing
between the ejecta and companion's envelope. In each case presented
here, the overall evolution and flow morphology bears close
resemblance to that of a supersonic flow past a sphere
\cite{hayes+04}. In that case, a bow shock forms at some distance from
the object, and the object becomes engulfed in a shocked gas. Provided
the fluid viscosity is sufficiently low, as is the case in the
situation considered here, a boundary layer forms close to the surface
of the object. This boundary layer eventually separates from the body,
and the flow recirculates and converges further downstream from the
object forming a turbulent wake \cite{vandyke82}.
\subsection{Binary system with main sequence companion MS38} \label{s:ms38results}
\begin{figure*}
\adjustThreePanels\ignorespaces
\begin{center}
    \begin{tabular}{ccc}
        \includegraphics{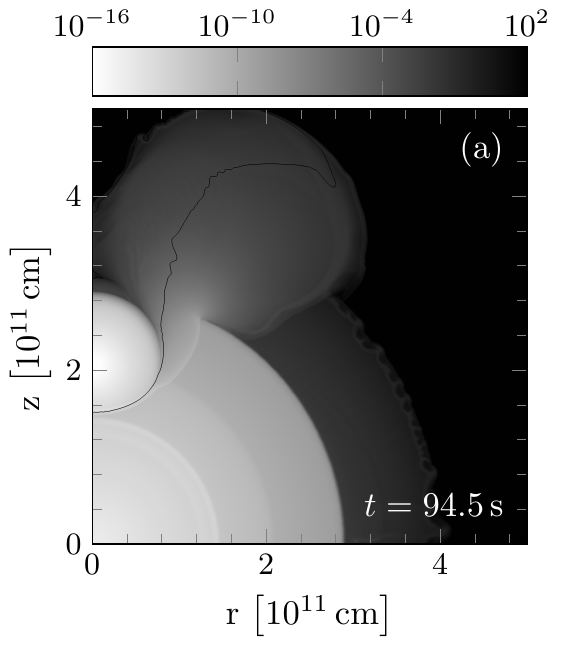}
        % \tikzImageTopbar{figs_publish/Fig3a_w7_ms38_logd_t_94p5s}{blackwhite_r}*[a][][white]*\ignorespaces
        &
        \includegraphics{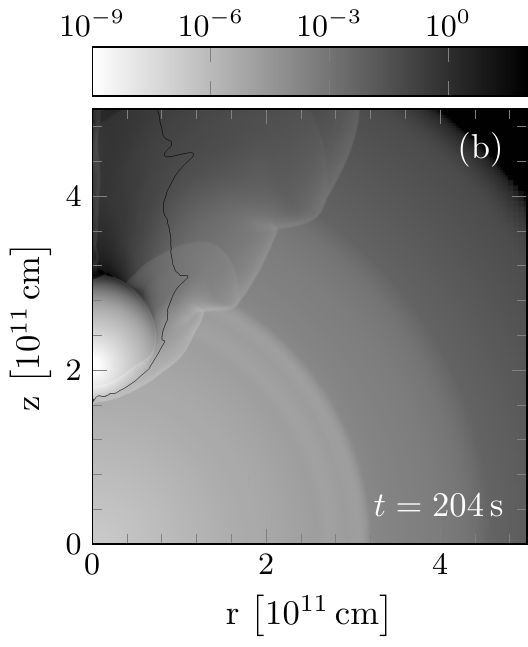}
        % \tikzImageTopbar{figs_publish/Fig3b_w7_ms38_logd_t_203p5s}{blackwhite_r}*[b][][white]*\ignorespaces
        &
        \includegraphics{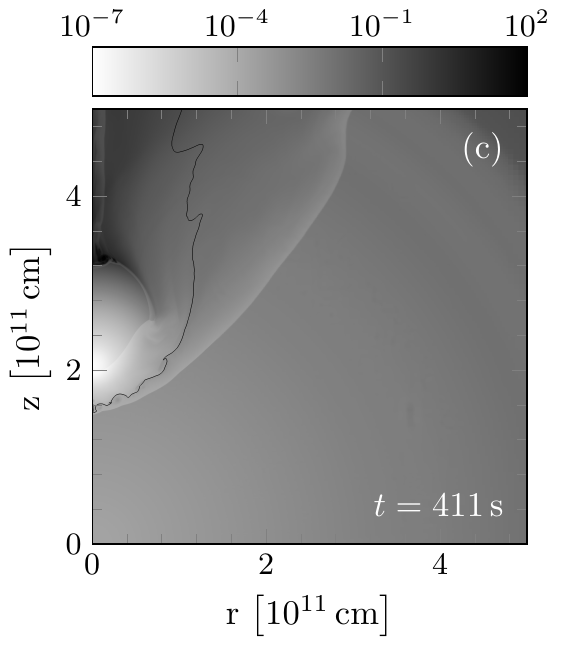}
        % \tikzImageTopbar{figs_publish/Fig3c_w7_ms38_logd_t_411p2s}{blackwhite_r}*[c][][white]*\ignorespaces
        %
    \end{tabular}
%
    % \blackLink{cb:mergerMorph0812Vert}
    % \includegraphics{figures/fig5colorbar.pdf}
% 
    \caption
    {Pseudocolor maps of density distribution in the MS38 model. (a)
      $t = 94$ s; (b) $t$ = 203 s; (c) $t$ = 411 s. In each panel, a
      contour line denotes the position of the contact discontinuity
      separating the ejecta from the stellar companion material. Note
      that the density scale changes between the panels. See text for
      details.  \label{fig:ms38_morph} }
\end{center}
\end{figure*}

Fig.\ \ref{fig:ms38_morph}
shows details of the interaction between the supernova ejecta and the
MS companion in the MS38 model. Fig.\ \ref{fig:ms38_morph}(a) shows
the density structure about a minute after the simulation has
started. At this time, a pair of shocks exist within the
supernova-companion interaction region. The first shock is the
original supernova shock transmitted into the stellar companion, which
by this time has penetrated around one third of the stellar radius
toward the companion's center. The second shock is the reflected
shock, which formed when the original supernova shock hit the steep
density gradient of the outer layers of the stellar companion. This
shock moves into the supernova ejecta. The two post-shock regions are
separated by a contact discontinuity, which delineates the shocked
supernova ejecta from the shocked companion material. The contact
discontinuity is depicted in Fig.\ \ref{fig:ms38_morph}(a) with a
solid line starting at $(r,z) \approx (0,1.5 \times 10^{11})$ cm. The
contact line is visibly flattened near $r = 0$ cm, indicating the
compression of the companion's envelope due to interaction with the
supernova ejecta.

The structure is somewhat more complex in the region behind the
companion. At the time of Fig.\ \ref{fig:ms38_morph}(a), the
companion is already completely engulfed in the shocked material. The
material stripped from the companion begins to accumulate in the
region behind ($z > 2.9 \times 10^{11}$ cm) the companion, which is
customarily identified as the ``ejecta hole.'' Further away from the
symmetry axis, the stripped material is separated from the ejecta by
the contact discontinuity (the solid line passing near $(r,z) = (1.2
\times 10^{11},4 \times 10^{11})$ cm). The contact discontinuity is
located inside a large, bubble-like region created when the high
pressure, doubly-shocked material (overrun first by the main supernova
shock and then by the reflected shock) provided additional pressure
support for the main supernova shock. Thus, this region is bounded
from the outside by the supernova shock.

The shock transmitted into the companion's interior weakens around the
progenitor's perimeter as the contribution of the ram pressure of the
supernova shock decreases. Similarly, the reflected shock becomes
gradually weaker, although it remains strong enough to move laterally
into the rapidly expanding, low-density ejecta (visible as a density
jump located near $(r,z) = (2.3 \times 10^{11},2.6 \times 10^{11})$
cm). The lower part of the expanding supernova remains unaffected by
the supernova-companion interaction. The outer edge of the expanding
structure is composed of a layer bounded by the supernova shock on the
outside and by a reverse supernova shock on the inside. The contact
discontinuity separating the shocked ambient medium material from the
shocked ejecta displays a complicated structure due to Rayleigh-Taylor
instability (RTI).

As the time progresses, the transmitted shock continues to move
through the stellar companion, and by $t = 204$ s it penetrates to
about two-thirds of stellar radius towards the center (see
Fig.\ \ref{fig:ms38_morph}(b)). The contact surface separating the
shocked stellar envelope from the shocked supernova ejecta begins to
show the first signs of Kelvin-Helmholtz instability (KHI) near $(r,z)
= (0,1.7 \times 10^{11})$ cm. This is due to progressively increasing
shear as the shocked ejecta flows around the companion star. The
region bounded by the transmitted and reflected shock now contains a
significant amount of material heated to temperatures around $5 \times
10^{7}$ K, which has been predicted to produce a significant amount of
X-ray emission \citep{Kasen+10}. We should point out that much higher
temperatures ($9 \times 10^{9}$\,K) are observed in the post-shock
region of the supernova shock. However, it is not expected that this
material is going to produce any significant amount of emission as the
amount of mass involved is small. At the time shown in
Fig.\ \ref{fig:ms38_morph}(b), the reflected shock continues to move
into the ejecta and its front is composited of three distinct
segments. This rather unusual shape of the shock front is due to the
density stratification of the ejecta. Another new flow feature visible
at this time is a conical shock located near the center of the
low-density region behind the companion. This shock is relatively weak
compared to other shocks present in this problem, and it would likely
be weaker in simulations that do not assume symmetry.

Fig.\ \ref{fig:ms38_morph}(c) shows the structure of the interaction
region at $t = 411$\,s. By this time the transmitted shock has reached
the stellar companion's core. The Kelvin-Helmholtz instability at the
interface separating the supernova ejecta from the stellar material is
now well developed, while the temperatures decreased slightly to $2-3
\times 10^{7}$\,K. The companion's layers and the ejecta shocked and
compressed at the beginning of the interaction process, which includes
the region associated with KHI, have now begun to expand with
velocities of up to 500 km~s$^{-1}$. Additionally, further along the
interface, as the shear increases and the compression decreases, the
outer layers of the companion are mixed with the ejecta and carried
away downstream with the flow. The stripped material is advected
around and past the companion, continuing to fill the low-density
region behind the companion.
\subsection{Binary system with subgiant companion SG} \label{s:sgresults}
\begin{figure*}
\adjustThreePanels\ignorespaces
\begin{center}
    \begin{tabular}{ccc}
        \includegraphics{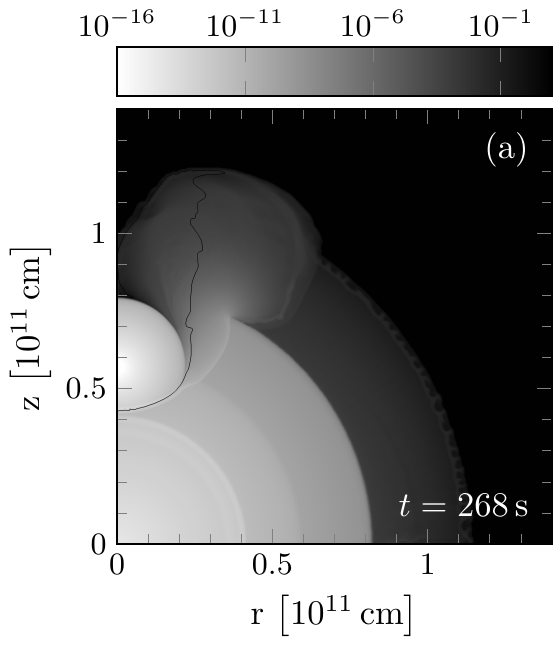}
        % \tikzImageTopbar{figs_publish/Fig4a_w7_sg_logd_t_268s}{blackwhite_r}*[a][][white]*\ignorespaces
        &
        \includegraphics{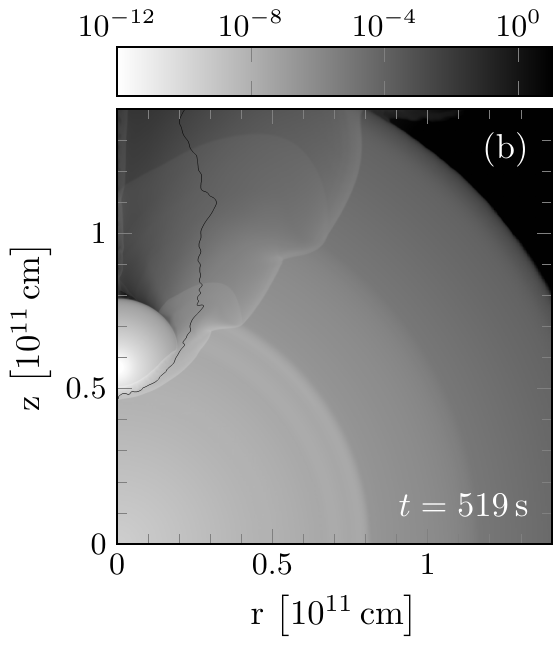}
        % \tikzImageTopbar{figs_publish/Fig4b_w7_sg_logd_t_519s}{blackwhite_r}*[b][][white]*\ignorespaces
        &
        \includegraphics{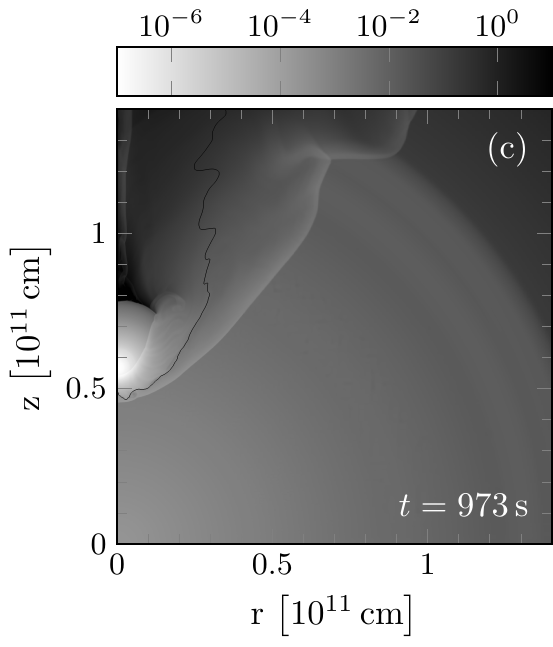}
        % \tikzImageTopbar{figs_publish/Fig4c_w7_sg_logd_t_973s}{blackwhite_r}*[c][][white]*\ignorespaces
        %
    \end{tabular}
    \caption
    {Pseudocolor maps of density distribution in the SG model. (a) $t
      = 268$ s; (b) $t = 519$ s; (c) $t = 973$ s. In each panel, a
      contour line denotes the position of the contact discontinuity
      separating the ejecta from the stellar companion material. Note
      that the density scale changes between the panels. See text for
      details.  \label{fig:sg_morph} }
\end{center}
\end{figure*}

The evolution in the case of the SG companion proceeds qualitatively
in a very similar matter as the main-sequence companion discussed in
Section \ref{s:ms38results}. The noticeable difference at early times
($t=268$\,s, Fig.\ \ref{fig:sg_morph}(a))
is that the hemisphere of the companion facing the ejecta is visibly
more compressed due to the more extended envelope of the SG. The
incoming ejecta is thus able to impart greater pressure upon the more
flat surface of the companion. (The geometrical effects, such as the
differing orbital distance, do not play a part here. This is because
in our models the secondary fills its Roche lobe and the companion
occupies nearly the same solid angle as seen from the explosion
center.) The transmitted shock is thus able to compress relatively
larger portion of the envelope. This increase in the efficiency of the
interaction can be seen by comparing the location of a kink at the
surface of the transmitted shock (visible near $(r,z) = (0.7 \times
10^{11},2.2 \times 10^{11})$ cm and $(r,z) = (0.2 \times 10^{11},0.62
\times 10^{11})$ in Fig.\ \ref{fig:ms38_morph}(b) and
Fig.\ \ref{fig:sg_morph}(b), respectively.) The kink is associated
with a point where the ejecta is no longer able to drive a strong
shock into the envelope. The more efficient energy transfer from the
ejecta into the envelope also enhances the process of stripping of the
envelope material from the companion, when compared to the more
compact main-sequence companion. At the time when the transmitted
shock has reached the companion's center ($t=973$\,s in
Fig.\ \ref{fig:sg_morph}(c)), and in contrast to the main-sequence
model (cf.\ \ref{fig:ms38_morph}(c)), almost the entire envelope has
been overrun by the transmitted shock. Furthermore, KHI is visibly
less active along the contact discontinuity separating the shocked
envelope and the shocked ejecta. This is because the density contrast
across the contact discontinuity is lower than in the case of the more
compact main-sequence companion (and the velocity shear is weaker).
\subsection{Binary system with red giant companion RG319} \label{s:rg319results}
\begin{figure*}
\adjustThreePanels\ignorespaces
\begin{center}
    \begin{tabular}{ccc}
        \includegraphics{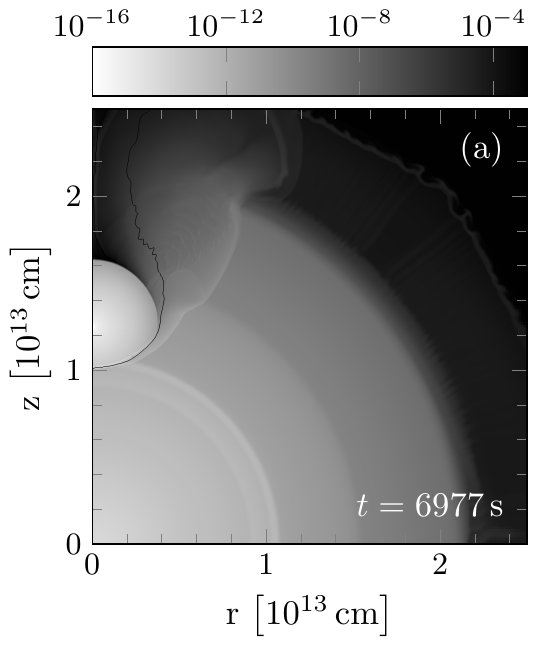}
        % \tikzImageTopbar{figs_publish/Fig5a_w7_rg319_logd_t_6977s}{blackwhite_r}*[a][][white]*\ignorespaces
        &
        \includegraphics{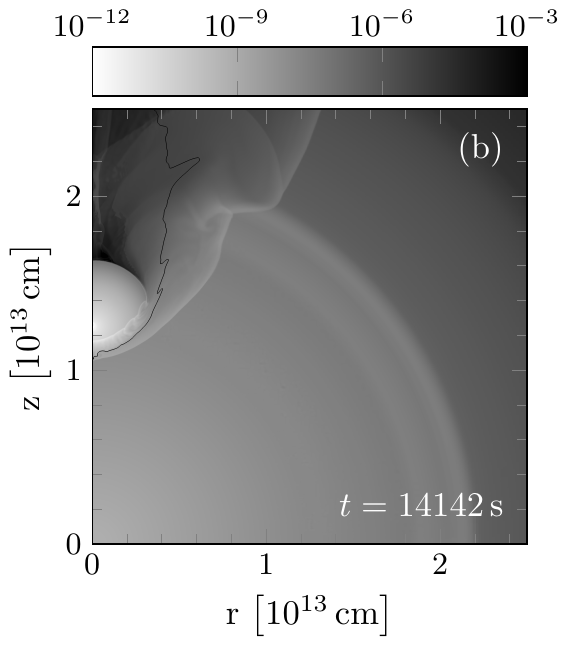}
        % \tikzImageTopbar{figs_publish/Fig5b_w7_rg319_logd_t_14142s}{blackwhite_r}*[b][][white]*\ignorespaces
        &
        \includegraphics{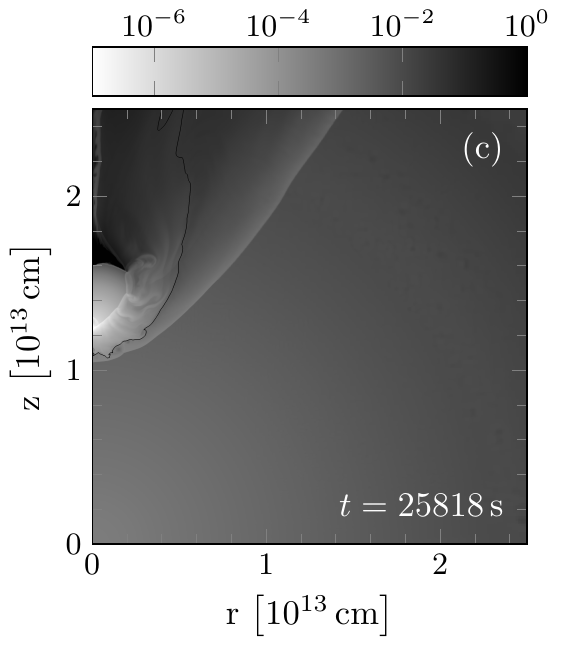}
        % \tikzImageTopbar{figs_publish/Fig5c_w7_rg319_logd_t_25818s}{blackwhite_r}*[c][][white]*\ignorespaces
        %
    \end{tabular}
    \caption
    {Pseudocolor maps of density distribution in the RG319 model. (a)
      $t = 6977$ s; (b) $t = 14142$ s; (c) $t = 25818$ s. In each
      panel, a contour line denotes the position of the contact
      discontinuity separating the ejecta from the stellar companion
      material. Note that the density scale changes between the
      panels. See text for details.  \label{fig:sy319_morph} }
\end{center}
\end{figure*}

Apart from the difference in temporal and spatial scales, the
evolution in the case of the RG319 binary proceeds in a similar
fashion to that in the SG system described in Section
\ref{s:sgresults}. For example, the compression of the companion's
envelope at early times (Fig.\ \ref{fig:sy319_morph}(a))
is only marginally stronger than the compression seen in the SG
case. Also, similar to the SG case, the KHI does not grow because of
the lower density contrast across the contact discontinuity and the
weaker shear (Fig.\ \ref{fig:sy319_morph}(b)). Finally, because the
envelope has a much lower density than in the SG case, the variations
in the ram pressure provided by the ejecta are now more easily
communicated to the reflected shock. Eventually this leads to the
``segmentation'' of the reflected shock front
(Fig.\ \ref{fig:sy319_morph}(c)).
\section{Discussion}\label{s:discussion}
\subsection{Structure of the ejecta hole}\label{ss:colDens}
\begin{figure*}
\adjustThreePanels\ignorespaces
\begin{center}
    \begin{tabular}{ccc}
        \includegraphics{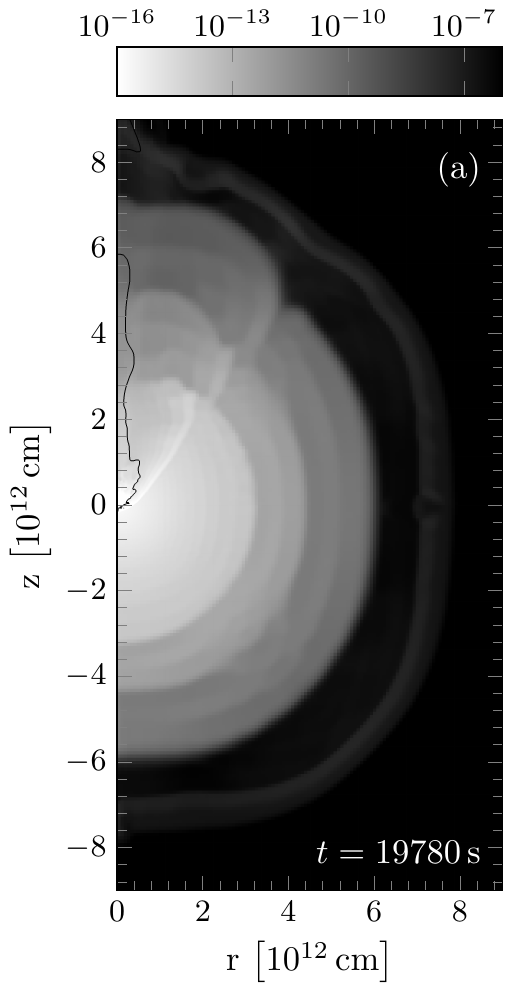}
        % \tikzImageFullDomain{figs_publish/Fig7a_w7_ms38_logd_t_19780s}{blackwhite_r}*[a][][white]*\ignorespaces
        &
        \includegraphics{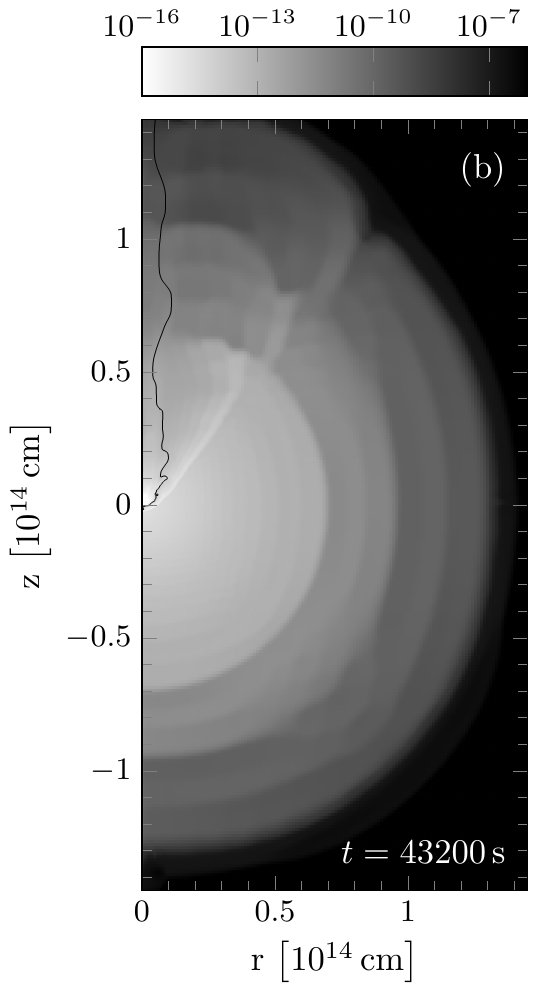}
        % \tikzImageFullDomain{figs_publish/Fig7b_w7_sg_logd_t_43200s}{blackwhite_r}*[b][][white]*\ignorespaces
        &
        \includegraphics{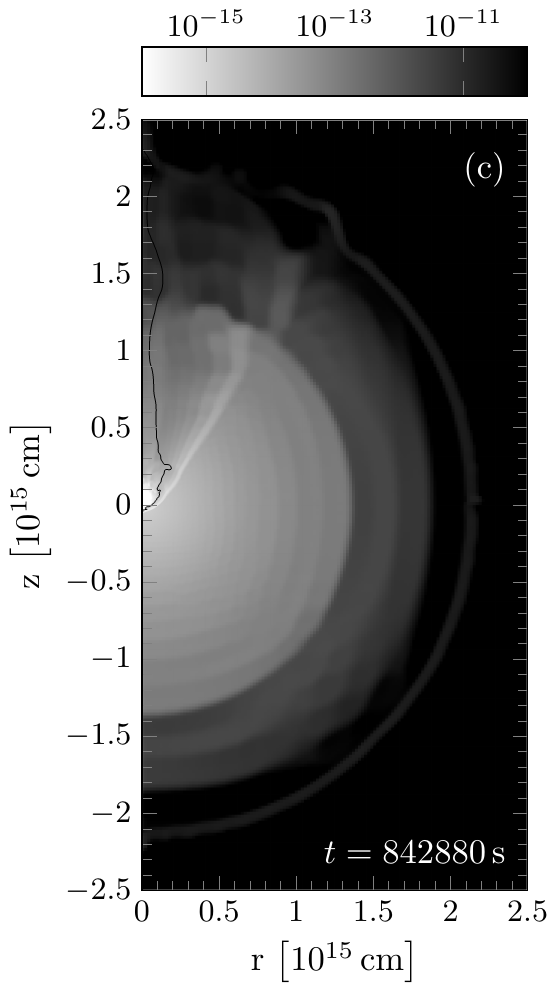}
        % \tikzImageFullDomain{figs_publish/Fig7c_w7_rg319_logd_t_842880s}{blackwhite_r}*[c][][white]*\ignorespaces
        %
    \end{tabular}
%
    % \blackLink{cb:mergerMorph0812Vert}
    % \includegraphics{figures/fig5colorbar.pdf}
% 
    \caption
    {Pseudocolor maps of density distribution in a subset of our
      binary system models at the final simulated times. (a) MS38 ($t
      = 19780$ s); (b) SG ($t = 43200$ s); (c) RG319 ($t = 842880$
      s). In each panel, a contour line denotes the position of the
      contact discontinuity separating the ejecta from the stellar
      companion material. Note that the density scale and spatial
      scale changes between the panels. See text for
      details. \label{fig:morph_final} }
\end{center}
\end{figure*}

Fig.\ \ref{fig:morph_final}
shows the structure of our supernova-companion interaction models at
the final simulated times. The ejecta hole is visible in each figure
panel in the upper half of the computational domain, extending from
the symmetry axis. The surface of the conical region is bounded by the
reflected (bow) shock, which appears ``segmented'' due to
stratification of the ejecta (cf.\ Section \ref{s:ms38results}). The
innermost regions of the hole are filled with the material stripped
from the companion, and the solid line marks the contact discontinuity
separating the companion's material from the supernova ejecta. The
abundance of the companion's material in the hole gradually decreases
with radius from the companion toward the supernova shock front, and
is completely mixed with the ejecta at the shock front. Although much
of the companion's material is confined near the center line of the
hole, small amounts of companion material can be found throughout the
hole region. Trace amounts ($\sim 1 \times 10^{-4}$) of the
companion's material can also be found further away from the hole's
center line along the surface of the supernova. This contamination
occurred during the early stages of the interaction and affects a
region extending 15-30 degrees outside the bow shock.

The geometry of the hole has been discussed in detail by a number of
authors. \citet{Marietta+00} originally reported the ejecta hole to
have an opening angle between 30 and 40 degrees depending on the
ejecta velocity (supernova explosion energy), with narrower holes
produced by ejecta with higher velocities. These results appeared
independent of the companion type (MS star, SG, and RG)
\citet{Pakmor+08} found the holes with opening angles about 45 degrees
for their series of main-sequence companions. Later on,
\citet{Garcia-Senz+12} reported holes with opening angles of 40
degrees for their single MS model considered (independent
of the model dimensionality or whether the orbital rotation was
accounted for or not). Around the same time, \citet{Pan+12a} reported
results of more than twenty models of binaries containing MS, helium,
and RG companions. In their study, the hole opening angle varied
between 30 degrees for the compact helium companion to up to 50
degrees for their RG companion model star.

\begin{figure*}
%
%\adjustTwoPanels
\ignorespaces
\begin{center}
    \begin{tabular}{cc}
        \includegraphics[width=0.45\textwidth]{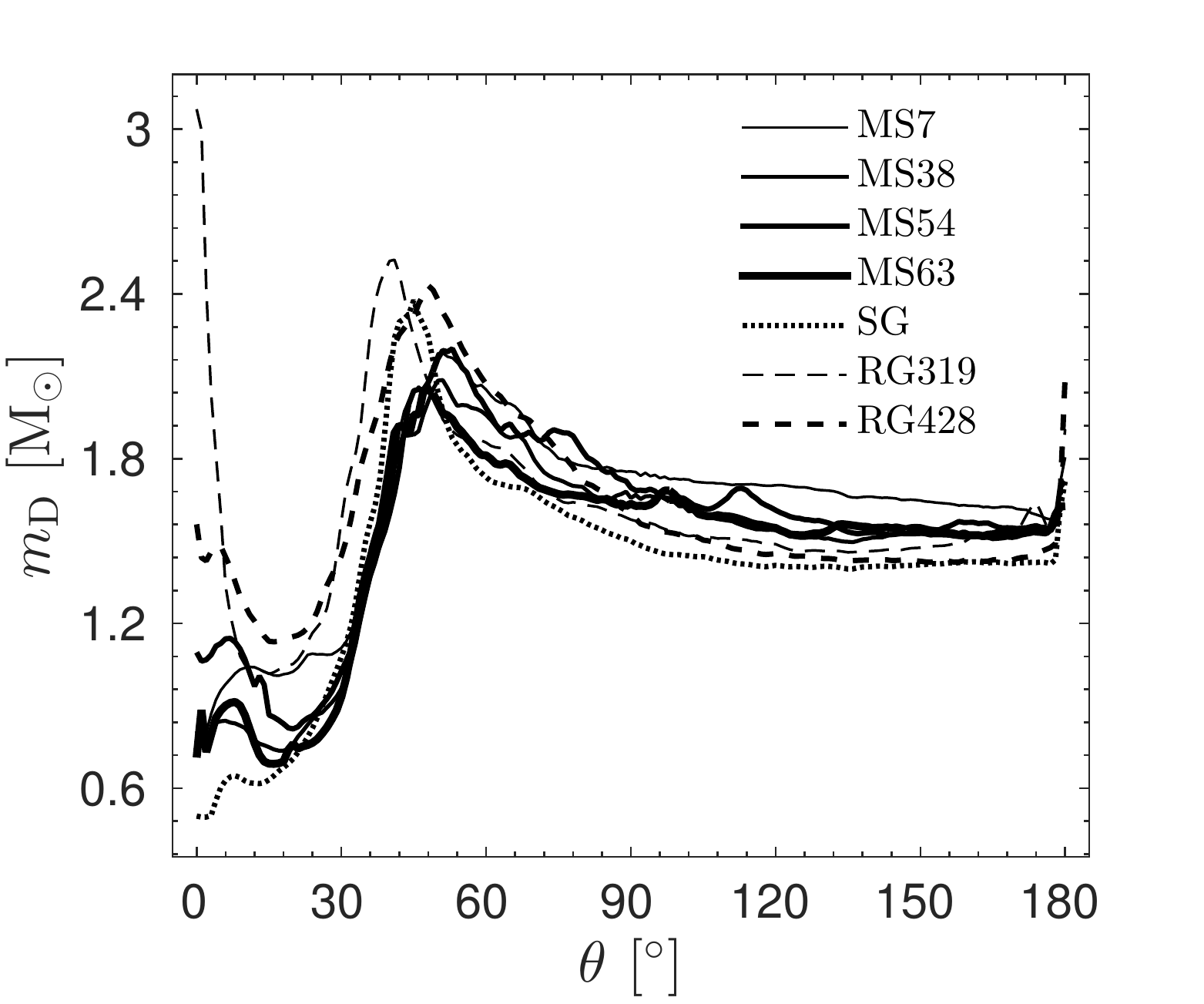}
        &
        \includegraphics[width=0.45\textwidth]{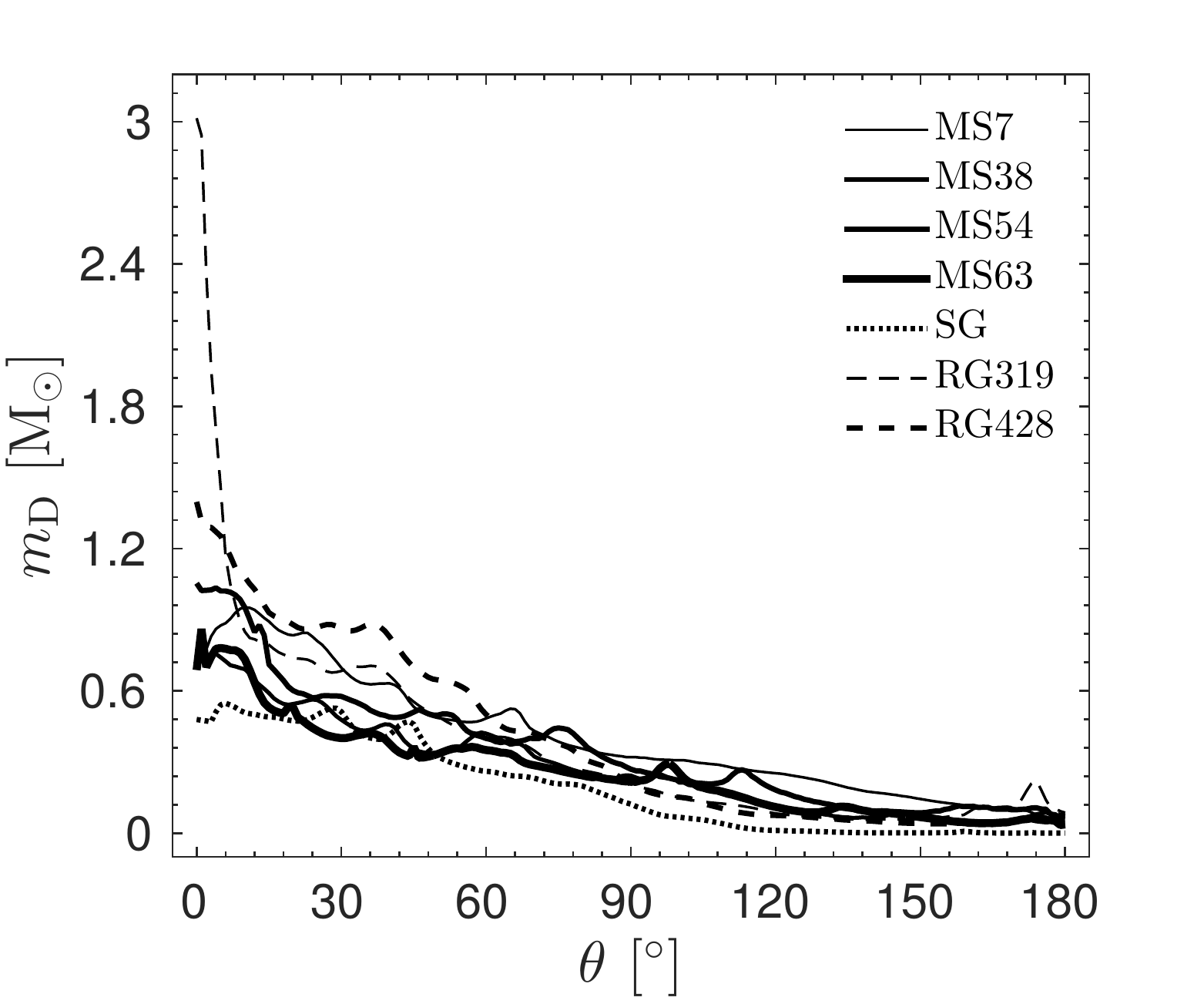}
    \end{tabular}
    \caption
    { Angular distribution of the equivalent surface mass,
      $m_{\mathrm{D}}$, at the final simulated times. The mass is
      accumulated along radial rays originating from the center of the
      companion star. The angle of 0\degrees\ corresponds to the ray
      aligned with the symmetry axis pointing in $z > 0$
      direction. (left panel) The unbound stellar material and the
      supernova ejecta material. (right panel) The unbound stellar
      material only. The mass distributions are shown with solid,
      dotted, and dashed lines for the MS, SG, and RG binary systems,
      respectively; see text for
      details.\label{fig:normalizedColumnDensities} }
\end{center}
\end{figure*}

Our results are largely consistent with the previous findings. To
illustrate the overall density distribution in the models, we
calculated an equivalent surface mass, $m_{\mathrm{D}}$. This quantity
has been obtained along radial rays originating at the symmetry axis
at the height corresponding to location of the maximum model density
(the central density of the companion). Starting from the origin of
the so-defined coordinate system, we divided rays in a number of small
radial segments. The beginning and the end of each segment were then
used to define the inner and outer radius, respectively, of spherical
shells. The mass of each shell was obtained by multiplying its volume
by the density interpolated at the point located at its average
radius. In the last step, the equivalent surface mass was obtained by
summing up mass contributions from all the shells except for the
material bounded to the companion. In other words, $m_\mathrm{D}$
measures the total mass of a spherically symmetric configuration whose
radial density profile is identical with the density profile measured
at a given time along rays originating from the approximate companion
center. The left panel in Fig.\ \ref{fig:normalizedColumnDensities}
shows the dependence of the equivalent surface mass as a function of
the polar angle for both the (unbound) companion material and the
supernova ejecta. In the majority of the models, with the exception of
the red giant RG319 binary system, the density reaches the maximum
between 40 and 50 degrees away from the symmetry axis. This angle
defines the half-opening angle of the ejecta and the location of the
edge of the hole. The density gradually decreases from the edge into
the region outside the hole. The inner part of the hole's edge is
relatively well-defined with the density rapidly decreasing, while the
density begins to increase inside the innermost 10--15 degrees. This
increase in density near the center line of the hole is due to
accumulation of mass stripped from the companion. This is clearly
visible by comparing the distribution of the equivalent surface mass
for both the unbound stellar material and the supernova ejecta (left
panel in left panel in Fig.\ \ref{fig:normalizedColumnDensities}) to
that of the unbound stellar material only (right panel in
Fig.\ \ref{fig:normalizedColumnDensities}).

We found the greatest accumulation of stripped material in RG models
that have the most extended, weakly bound envelopes, while we observed
the least accumulation in the case of the compact SG. The amount of
stripped mass appears approximately linearly decreasing from the
symmetry axis up to at least 120\degrees\ away from the center line of
the hole. No stripped material is observed at greater angular
distances from the hole in the case of the SG model, where in other
models the amount of stripped material observed in those regions does
not exceed 10 per cent of the typical values found near the hole's
center.
\subsection{Amount of stripped mass}\label{ss:strMass}
To estimate the amount of mass lost by a stellar companion, we
compared its initial mass (cf.\ Table \ref{t:simParams}) to the amount
of gravitationally bound stellar material found at the final
simulation time. Table \ref{t:quantities}
\begin{table}
 \centering
  % \begin{minipage}{40mm}
  \caption{Total amount of mass stripped from the companion, $\Delta_m$, and the fraction of each companion's initial mass that was stripped, $\Delta m/m_{\ast}$, at the final simulated times.\label{t:quantities}}
  \begin{tabular}{l c c}
    \hline
    Model       & $\Delta m$   & $\Delta m/m_{\ast}$ \\
    designation & [M$_{\odot}$] &                       \\
    \hline
    MS7   & 0.37                        & 0.24               \\
    MS38  & 0.25                        & 0.22               \\
    MS54  & 0.32                        & 0.26               \\
    MS63  & 0.24                        & 0.21               \\
    SG    & 0.17                        & 0.11               \\
    RG319 & 0.28                        & 0.41               \\
    RG428 & 0.33                        & 0.44               \\
    \hline
  \end{tabular}
  % \end{minipage}
\end{table}
shows the final masses of and relative mass lost by our companion
model stars. For the given supernova explosion (energy, ejecta mass)
and orbital (distance, Roche-filling factor) parameters, the
gravitational binding energy of the companion star is the key factor
that determines how much mass the companion will lose during the
interaction with the supernova.

The mass stripping process was the most efficient in the case of RG
model binaries, in which case both red giant companions lost their
entire envelopes. The typical amount of mass stripped in the case of
MS companions was slightly over 20 per cent, while only
about 10 per cent of the companion mass was lost in the case of the
SG. The observed amounts of stripped mass are not surprising
given that the envelope stripping process sensitively depends on the
binding energy of the envelope. In addition, the mass loss occurs
primarily during the first hour of the interaction in the case of SG
and MS companions, and between half a day and one day in the case of
RG companions.

The relative amount of stripped mass in the case of our MS binary
models varies between 21 and 26 per cent. This is typically higher
than the amounts of stripped mass reported by \cite{Marietta+00}
($\approx$15 per cent for their HCV model), \cite{Pan+12a}
($\approx$16 per cent in their 2D, non-rotating, Roche-lobe-filling
MS-2D-Nr model), and \cite{Liu+12} (between 6 and 24 per cent across
their MS models). The mass loss found in our MS models approximately
follows the power-law relation between stripped mass and orbital
separation obtained by \cite{Pakmor+08}. However, the obtained mass
loss is 30--60 per cent greater than estimate given by equation 4 in
\cite{Pakmor+08}, even once corrected for the differences in supernova
explosion energies (using equation 2 in \cite{Pakmor+08}). Our results
are marginally consistent with the results presented by
\cite{Garcia-Senz+12}, who reported a mass loss of around 10 per cent.

In the case of a SG companion, \cite{Marietta+00} reported a
mass loss on the order of 15 per cent for their HCVL model, which
favorably compares to the estimated mass loss found in our SG
model. These estimated amounts of mass lost by SG companions exceed by
a factor of about 3 the mass loss reported by \cite{Pan+12a} for their
compact helium companion model. Finally, the groups who studied the
interaction between the supernova and the RG companions
\citep{Marietta+00,Pan+12a} uniformly found the companion stars
completely stripped of their envelopes, as is the case in our study.

We note that the results of \cite{Pakmor+08}, who reported a mass loss
of up to 5 per cent in their set of MS models, remains an outlier in
this series of studies of binaries with MS companions. The difference
in mass loss predicted in that study appears most likely as a result
of their use of companion models that were not computed
self-consistently in binary system evolution. (The later followup
study by \cite{Liu+12} who used companion models computed with a
binary stellar evolution code produced mass loss estimates similar to
those obtained by other groups.)
\subsection{Kinematic properties of stripped companion material}\label{ss:velDist}
Because no substantial amounts of hydrogen or helium are expected to
exist in the supernova ejecta, detection of hydrogen or helium lines
in the Type Ia supernova spectra would provide strong evidence for the
presence of a non-degenerate companion in the system. Therefore, such
detection would not only confirm the single-degenerate scenario for
SNIa formation, but also allow to distinguish between the supernovae
originating in different channels. \cite{Marietta+00} were the first
to provide estimates of the amount and velocity distribution of
hydrogen and helium stripped from the non-degenerate companion
stars. Their results indicated that the typical velocity of the
stripped hydrogen, defined as the half-mass point of the mass
distribution in velocity space, was about 800 km~$^{-1}$ for their
compact MS companion model, nearly 900 km~s$^{-1}$ in the case of
their SG companion, and between 400 and 600 km~s$^{-1}$ for their RG
companions.

\begin{figure*}
\adjustThreePanels\ignorespaces
\begin{center}
    \begin{tabular}{ccc}
        \includegraphics[width=0.33\textwidth]{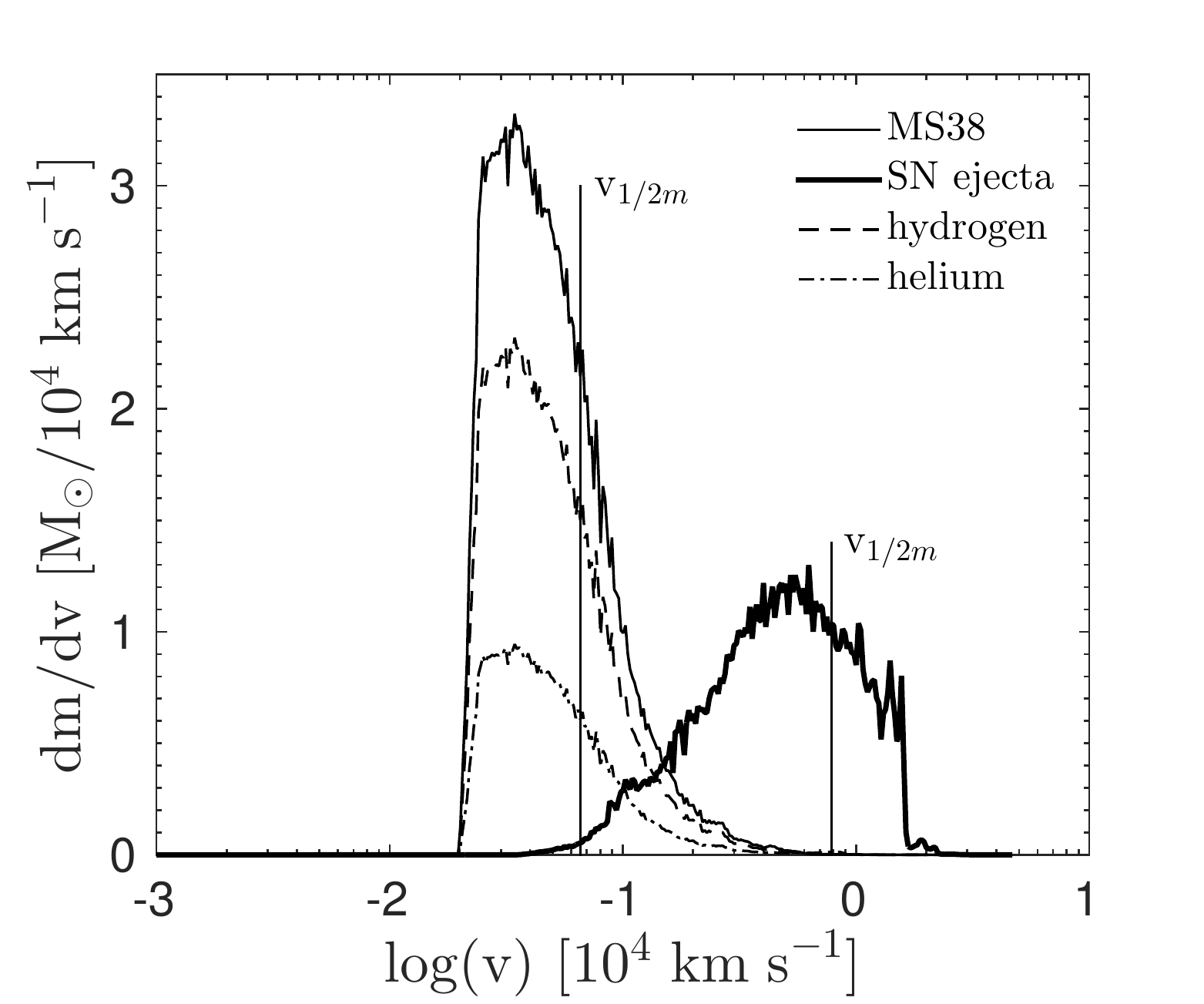}
        % \tikzImageFullDomain{figs_publish/Fig7a_w7_ms38_logd_t_19780s}{blackwhite_r}*[a][][white]*\ignorespaces
        &
        \includegraphics[width=0.33\textwidth]{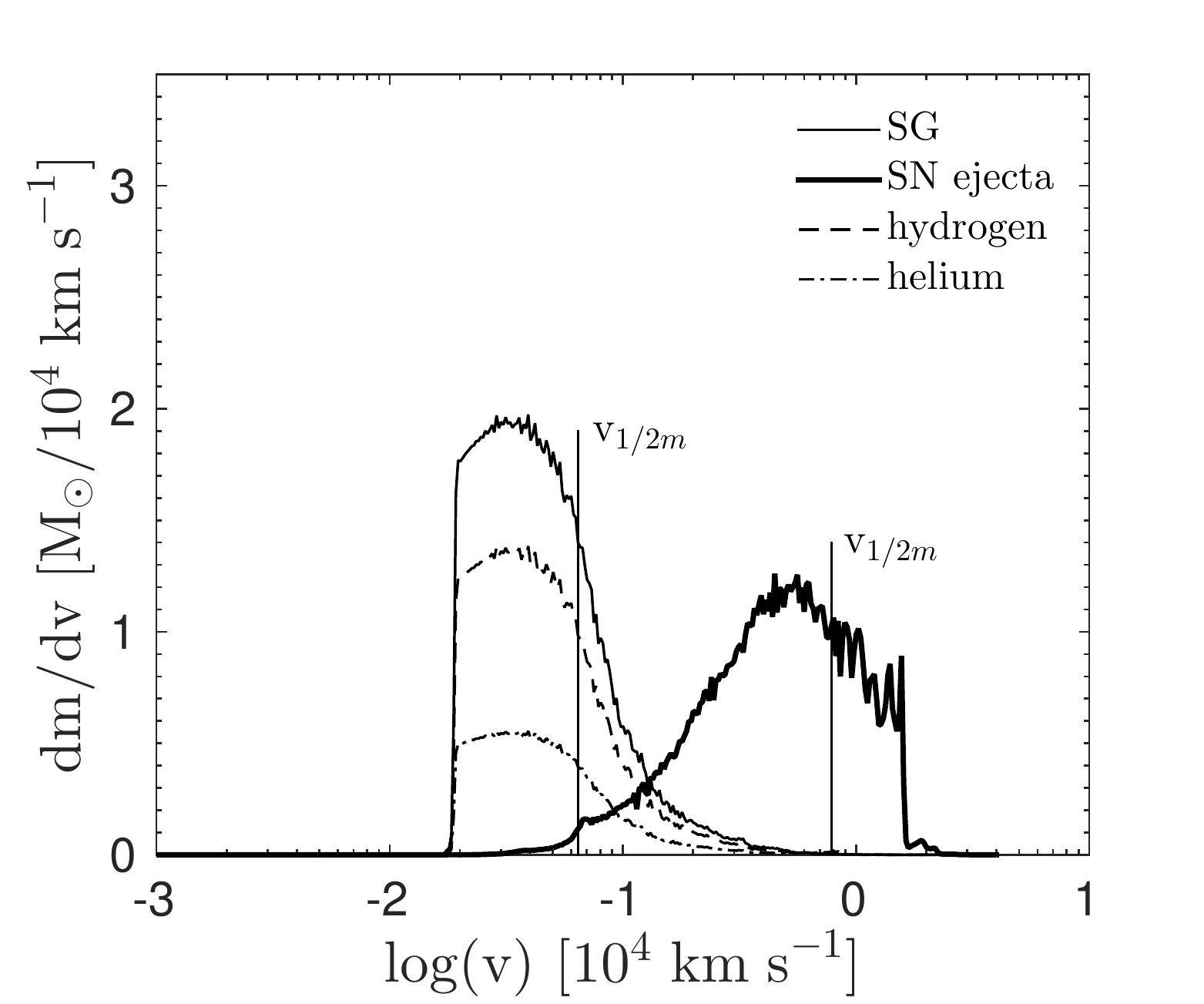}
        % \tikzImageFullDomain{figs_publish/Fig7b_w7_sg_logd_t_43200s}{blackwhite_r}*[b][][white]*\ignorespaces
        &
        \includegraphics[width=0.33\textwidth]{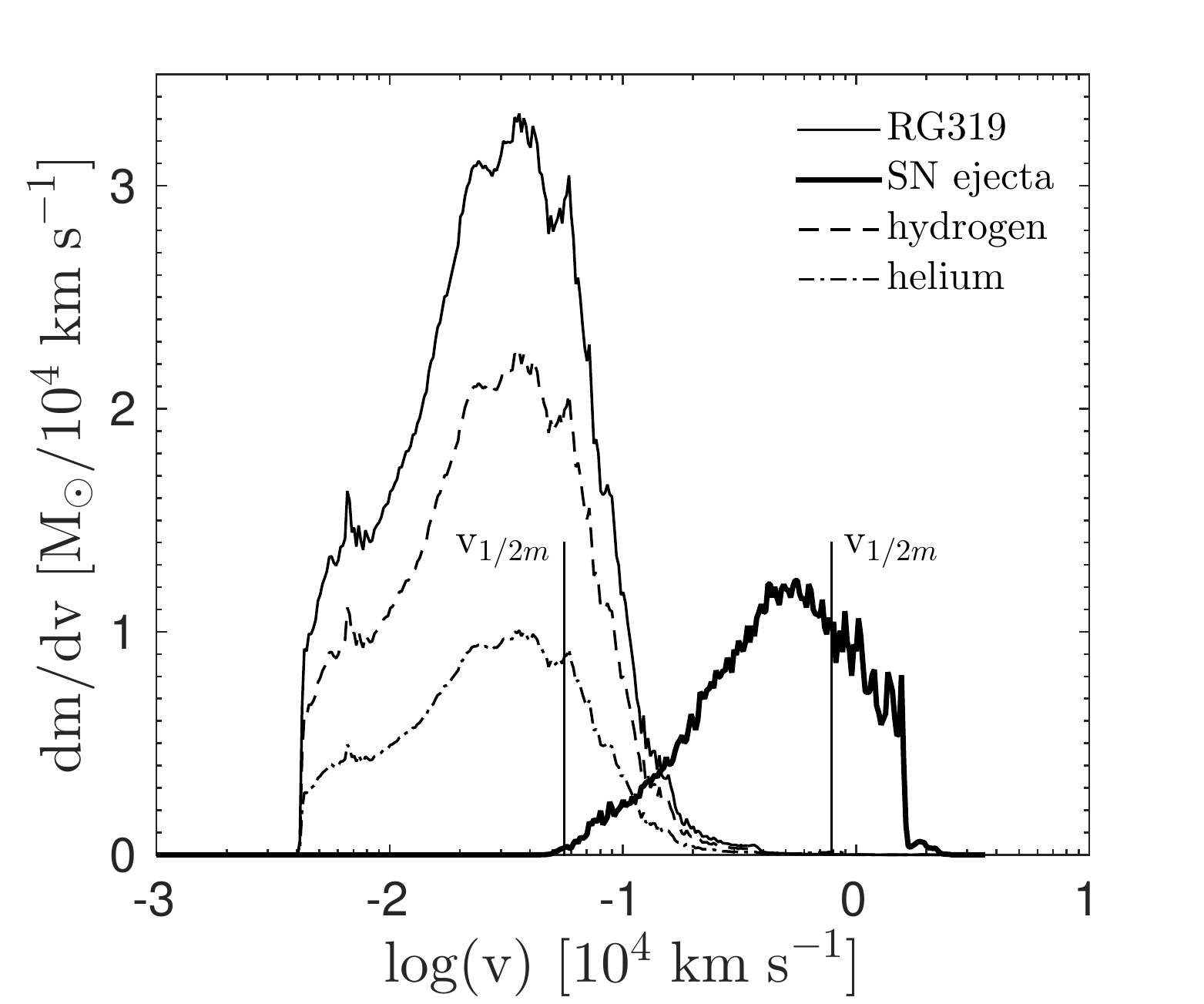}
        % \tikzImageFullDomain{figs_publish/Fig7c_w7_rg319_logd_t_842880s}{blackwhite_r}*[c][][white]*\ignorespaces
        %
    \end{tabular}
%
    % \blackLink{cb:mergerMorph0812Vert}
    % \includegraphics{figures/fig5colorbar.pdf}
% 
    \caption
    {Velocity distribution of the unbound companion material and the
      supernova ejecta for a subset of companion models at the final
      simulated times. (left panel) the main-sequence companion MS38;
      (middle panel) the subgiant SG; (right panel) the red giant
      RG319. For each model, the mass distribution is shown for
      hydrogen (dashed line), helium (dash-dotted line), the total
      stripped companion material (thin solid line), and the supernova
      ejecta (thick solid line). Short vertical lines indicate the
      average velocities of the stripped companion material (left
      vertical line) and the supernova ejecta (right vertical
      line). Note that the velocity is shown in log scale. See text
      for discussion.\label{fig:velocity_final} }
\end{center}
\end{figure*}

Fig.\ \ref{fig:velocity_final} (left-hand panel)
shows the mass distribution in velocity space in our MS38 binary
model. The velocity distribution of the ejecta material matches that
reported by \cite{Marietta+00} closely, as expected due to the use of
the same explosion model. The difference in the average velocities,
$v_{1/2m}$, can in part be explained by the supernova explosion energy
employed by \cite{Marietta+00} being approximately 7 per cent higher
than in our model. The other contributing factor, as we discuss below
in the context of the \cite{Pan+12a} work, might be lower mesh
resolution used in the \cite{Marietta+00} study.

The overall shape of the total stripped mass distribution, and also
hydrogen and helium distributions, found in our MS38 model is also
similar to those reported by \cite{Marietta+00}. We find higher
amounts of stripped material for a given velocity, which reflects the
fact that we find more stripped material in all of our
models. Additionally, as we mentioned earlier, the higher supernova
explosion energy used in \cite{Marietta+00} could be responsible for
their higher average velocities of the stripped material, $v_{1/2m}$
(823 km~s$^{-1}$ in their HCV model compared to our MS38 result of
approximately 657 km~s$^{-1}$). The only major qualitative difference is the
presence of a distinct low-velocity cutoff in the stripped mass in our
simulations. After careful examination of the data, we found that the
velocity associated with the cutoff in our results corresponds to the
velocity of the material that has zero binding energy. This implies
that at this particular moment during the evolution, the material with
near-zero binding energy is not in hydrostatic equilibrium, and
instead continues to expand away from the secondary. At later times,
however, it will fall back to the star. Because the velocity
distributions shown in \cite{Marietta+00} do not display such a
low-velocity cutoff, and evolutionary times in their corresponding
simulations are identical to ours, we conclude that the dynamics of
the material with near-zero binding energies differ.

Much of the same conclusions can be drawn from comparing the velocity
distribution of stripped material in our SG model (shown in the center
panel in Fig.\ \ref{fig:velocity_final}) to that of the HCVL model of
\cite{Marietta+00}. The average velocity distribution in our SG model
is approximately 642 km~s$^{-1}$, lower than the value reported in their work
(890 km~s$^{-1}$). Also, our velocity distributions indicate greater amounts
of stripped material than those reported by \cite{Marietta+00}
(cf.\ Table \ref{t:quantities}). Finally, in the case of our RG319
model, the average velocity of the stripped companion's material
(approximately 562 km~s$^{-1}$) compares favorably to that reported by
\cite{Marietta+00} in the case of their HALGOLa model (593
km~s$^{-1}$). However, the amount of mass stripped in our model
(approximately 0.3 M$_{\odot}$) is less than the amount reported in
their study (0.54 M$_{\odot}$).

\cite{Liu+12} presented analysis of velocity distributions of stripped
material only for one of their main-sequence models. Based on fig.\ 10
in their paper, we estimated the half-mass velocity of the stripped
material to be around 1400 km~s$^{-1}$. This is remarkably higher than
the average velocity reported by \cite{Marietta+00}, although both
groups used identical supernova explosion energies. It is also
substantially higher than indicated by the simulations of
\cite{Pakmor+08} (as reported by \cite{Liu+12}). One possibility for
this discrepancy could be a difference in the companion
models. However, our analysis of a similar in mass MS companion, MS54,
showed a half-mass velocity of about 600 km~s$^{-1}$. The results
presented by \cite{Garcia-Senz+12} do not allow for a quantitative
comparison, however the overall shape of the velocity distribution is
similar to that found in our MS38 main sequence model. Also, our
velocity distributions are higher by a factor of about 5 than those
reported in their work. However, it seems difficult to reconcile that
difference solely with their supernova model explosion energy being
only 20 per cent lower than that of our model supernova.

We observe much better qualitative agreement of velocity distributions
of stripped material between our results and those reported by
\cite{Pan+12a}. Although we cannot compare the actual values of the
velocity distributions due to differing units used to present the
results, the distributions presented by \cite{Pan+12a} show distinct
low-velocity cutoffs in the case of their MS and helium companions
while the velocity distribution extends to low velocities in the case
of their RG model system. The velocities corresponding to the maximum
mass distributions (not to be confused with the average velocity,
$v_{1/2m}$, discussed above) reported by \cite{Pan+12a} were, however,
substantially higher than those found in our models. More
specifically, the peak velocities of the stripped material were by
about 30 per cent, 60 per cent, and 40 per cent lower in the case of
our MS, SG, and red giant models, respectively. It is
unlikely that the lower supernova explosion energy used in our study
is responsible for these substantial differences in peak velocities of
stripped material. This is because the explosion energies used in the
two studies differs by only a few per cent. We conclude that either
the significantly lower mesh resolution or the particle-based analysis
method used by \cite{Pan+12a} are likely factors responsible for the
observed differences in the energetics of the stripped material.
\subsection{Stripped hydrogen-rich material at high velocities}\label{ss:velDistHigh}
\begin{figure}
  \centering
  \includegraphics[width=0.45\textwidth]{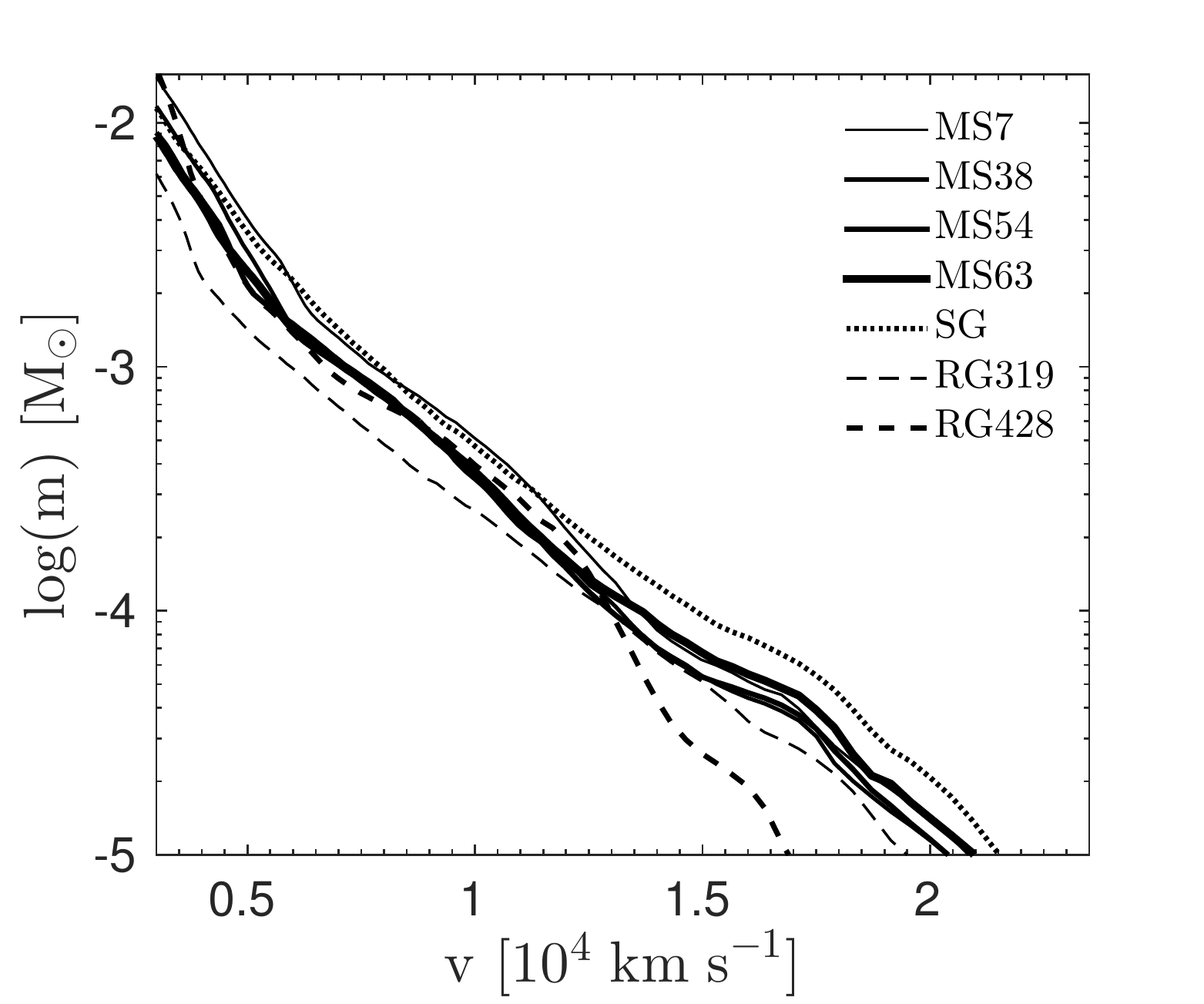}
    \caption{The amount of mass with velocities higher than a given
      value for unbound companion material at the final simulated
      times. The accumulated mass distributions are shown for MS, SG,
      and RG companions with solid, dotted, and dashed lines,
      respectively.. Note that the accumulated mass is shown in log
      scale; see text for details. \label{fig:velocity_high}}
\end{figure}

We conclude our discussion of kinematic properties of stripped
material by focusing on hydrogen within its high velocity tail. As
originally pointed out by \cite{Marietta+00}, the distribution of
hydrogen at high velocities is potentially an important
observationally accessible diagnostic allowing to discriminate between
various types of companion stars. Fig.\ \ref{fig:velocity_high}
shows the accumulated amount of stripped hydrogen mass with velocities
greater than the given velocity for our set of binary system
models. Several important characteristics of the hydrogen mass
velocity distributions can be noted. Despite the diversity of our main
sequence model sample, their hydrogen mass velocity distributions are
very similar. In addition, the distributions for all companion models
are similar for velocities below approximately 13\,000 km~s$^{-1}$. At
higher velocities, companions of different types fold into distinct
groups. For example, we can clearly distinguish between the SG model
and MS models, with the SG containing a greater amount of mass at a
given velocity. Both RG models, especially the most extended RG428
companion, produce the least amount of high velocity hydrogen, and
should be relatively easily distinguishable from other companion
types. The amount of hydrogen mass moving with the velocities above
15\,000 km~s$^{-1}$ in the SG model is by a factor almost 4 greater
than that of the RG428 model.

Our accumulated hydrogen masses as a function of velocity can be
compared to the results of \cite{Marietta+00} (see their
fig.\ 32). Although the overall run of the distributions appears very
similar between the two studies, the hydrogen mass distributions are
not as clearly separated in the \cite{Marietta+00} study than in our
models. There is potential confusion between various companion models
in their study at velocities around 19\,000 km~s$^{-1}$, and the
distributions cross over again at around 13\,000 km~s$^{-1}$. The
amount of hydrogen mass moving faster than 15\,000 km~s$^{-1}$ in the
\cite{Marietta+00} study differs by a factor of about 2.6 between
their models, which is somewhat less than the factor of almost 4 found
in the present work. It is conceivable that the differences in the
kinematics of the stripped hydrogen found in our simulations as
compared to those of \cite{Marietta+00} can be due to a number of
differences between the models such as model resolution, structure of
companion stars, and supernova explosion energy. Although we cannot
clearly identify the source of the differences, the observed
variations in the stripped hydrogen kinematics provide initial insight
into the uncertainties of the model predictions based on this kind of
diagnostic.
\subsection{Chemical contamination of companion}\label{ss:contamination}
It is conceivable that the supernova ejecta could deposit its material
onto the companion. Thus, the presence of lines of elements that are
not produced in low mass stars would provide evidence for the
interaction between the companion and the supernova. To quantify the
level of contamination of the companion by the ejecta, we calculated
the amount of $^{56}$Ni that was bound to the companion star at the
final simulated time. The amounts of companion-bound nickel were less
than 10$^{-8}$ \Msun\ in the case of the MS companions,
around $10^{-7}$ \Msun\ in the case of the SG, with only trace
amounts of $^{56}$Ni found in the case of RG companions.

The estimated levels of contamination are relatively small compared to
other studies. \cite{Marietta+00} found that approximately $1.3 \times
10^{-4}$ M$_{\odot}$ of iron-rich supernova material was bound to the
companion star in their MS model at the end of their simulation. Their
estimate was by a factor of about 10 higher than the approximately
$10^{-5}$ M$_{\odot}$ reported by \cite{Pan+12a} in the case of their
three-dimensional model of supernova interaction with the MS
companion. Additionally, the work by \cite{Pan+12a} is the only study
that considered contamination of post-MS companions. In particular,
they reported a relatively low, but still significantly higher than
found in our model, contamination of their RG binary companion by
10$^{-8}$ \Msun\ of $^{56}$Ni. We are not in a position to compare
contamination estimation in the case of our SG model to the helium
model of \cite{Pan+10} or \cite{Pan+12a}, as the two binary systems
and companions differ substantially.

Because the amounts of ejecta material contaminating the secondary
stars found in this work and in the previous studies are relatively
small, it is conceivable that these estimates are subject to numerical
effects and certain model assumptions. In particular, we note that in
our MS models and in the case of our SG model, the
ejecta material bound to the secondary is the least mixed in the
region close to the symmetry axis and on the hemisphere of the
secondary that faces the explosion site.

The degree of mixing of the ejecta material with the envelope material
of the companion gradually increases further away from the symmetry
axis. This is chiefly because the primary mechanism for mixing in the
interaction models is shear between the supernova ejecta and the
stellar material. The shear is the least near the symmetry axis, which
explains the relative purity of ejecta material bound to the
companion. The outcome would almost certainly be different in more
realistic models that would account for the orbital motion and the
spin of the companion, as considered by \cite{Pan+12a}. In particular,
these authors found a substantial degree of mixing of the ejecta and
stellar companion material in the case of their helium model
(cf.\ left panel of fig.\ 23 in \citep{Pan+12a}). Because the amounts
of ejecta material contaminating the companion are subject to strong
numerical effects, as discussed in Section \ref{ss:resolution},
predictions based on low-resolution models should be considered
preliminary. In addition, as originally noted by \cite{Marietta+00},
the truthful estimates of contamination levels of companions would
require simulations covering much longer evolutionary times to allow
for accretion of ejecta material originally located close to the
ejecta center and, therefore, moving at relatively low speeds.
\subsection{Properties of transient X-ray source}\label{ss:tempDist}
The supernova shock interaction with the companion serves not only to
heat the companion's outer layers, but also the ejecta itself due to
the formation and propagation of the reflected shock, as we described
in Section \ref{s:ms38results}. A semi-analytic model shown by
\cite{Kasen+10} predicts that the ejecta heated by the reflected shock
may produce a highly luminous burst ($L\approx 10^{44}$~erg s$^{-1}$)
of soft X-rays. This theoretically predicted emission would dominate
the supernova luminosity for up to a few days, depending on the
companion type, and be strongly angle-dependent with most radiation
emitted through the (least opaque) hole region carved out in the
ejecta by the companion. Due to the hole geometry, the external
observer may have a chance to see the resulting soft X-ray/UV flash of
radiation in about 10 per cent of Type Ia events (originating from a
single degenerate formation channel).

\begin{figure}
  \centering
  \includegraphics[width=0.45\textwidth]{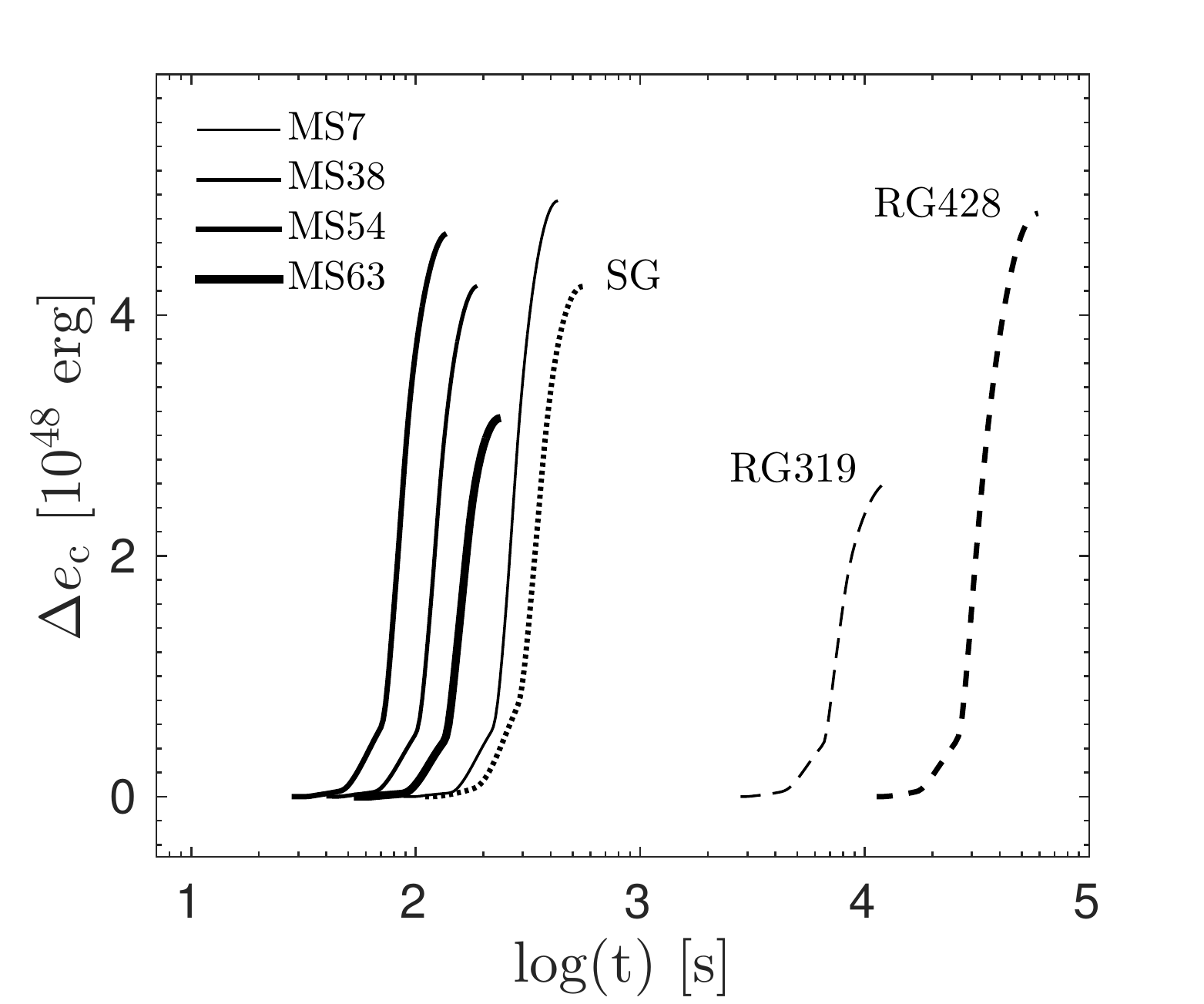}
    \caption{Change in internal energy of companion material in
      respect to the initial internal energy as a function of
      time. The energy change is shown with solid, dotted, and dashed
      lines for MS, SG, and RG models, respectively. The curves
      terminate at the times when the maximum energy change was
      recorded. Note that the time is shown in log scale; see text for
      details.\label{fig:eFnTime}}
\end{figure}

The shock-heating process of the stellar envelope in all of our model
binary systems is illustrated in Fig.\ \ref{fig:eFnTime}.
The evolution of the internal energy change (compared to the initial
internal energy) in every model is shown with a curve starting at the
initial simulation time and terminating at the time when the maximum
change of internal energy due to heating is recorded, $t = t_{h}$. We
do not show the data past that point in time, as the evolution of the
internal energy of the companion becomes affected by its expansion and
mixing of the envelope material with the ejecta. The time over which
the companion is heated chiefly depends on the companion's size. It
varies from as little as 80 seconds in the case of the MS54 companion
to as much as nearly half a day in the case of the RG428 model
giant. The estimated amount of energy stored during that time, along
with the energy stored in the ejecta, as discussed below, provides the
energy budget for the prompt radiation burst.

The contribution of the heated envelope component to the radiation
flash was not considered by \cite{Kasen+10}, who focused on the
evolution and emission of the ejecta.\footnote{As pointed out by
  \cite{Kasen+10}, the emission produced by this material has the best
  chance of being observed at early times. This is because, and in
  contrast to the shock-heated envelope, it remains unobscured by the
  dense ejecta when viewed at angles for which the opacity is
  sufficiently small. Such low opacities at early times can only be
  found for viewing angles inside the ejecta hole.} Furthermore, the
shocked ejecta and the shocked envelope material are subject to mixing
due to Kelvin-Helmholtz instability (cf.\ Section
\ref{s:results}). This mixed, hot plasma is advected around the
companion and lines the inner surface of the hole. Therefore, it will
also contribute to the overall burst emission. For completeness, we
would like to mention that both the unshocked ejecta and the unshocked
envelope will be irradiated by the corresponding shocked material
lying across the contact discontinuity. The irradiating components
will only be partially reflected with the majority of X-ray photons
down-scattered to lower energies \citep[see, e.g.,][]{Plewa95}.

Although we are not in a position to predict the actual contribution
to the emission from the companion (the mixed material or effects due
to irradiation), our simulation results provide information about the
amount of the energy deposited by shocks in the ejecta and the
envelope, along with their corresponding characteristic
temperatures. By comparing the characteristics of the shock-heated
envelope to those of the shock-heated ejecta, combined with the
semi-analytic predictions by \cite{Kasen+10}, we can provide upper
limits for the contribution of the shock-heated envelope material to
the flash emission. We note that the results of our hydro simulations
contain all of the information required to perform realistic radiation
transport calculations of the prompt emission. Such calculations are,
however, beyond the scope of the present study. Here we first consider
the ejecta material heated by the reflected shock, which is created
when the forward supernova shock penetrates into the companion's
envelope. Then we discuss the properties of the shocked stellar
envelope.

\begin{figure*}
%
%\adjustTwoPanels
\ignorespaces
\begin{center}
    \begin{tabular}{cc}
        \includegraphics[width=0.49\textwidth]{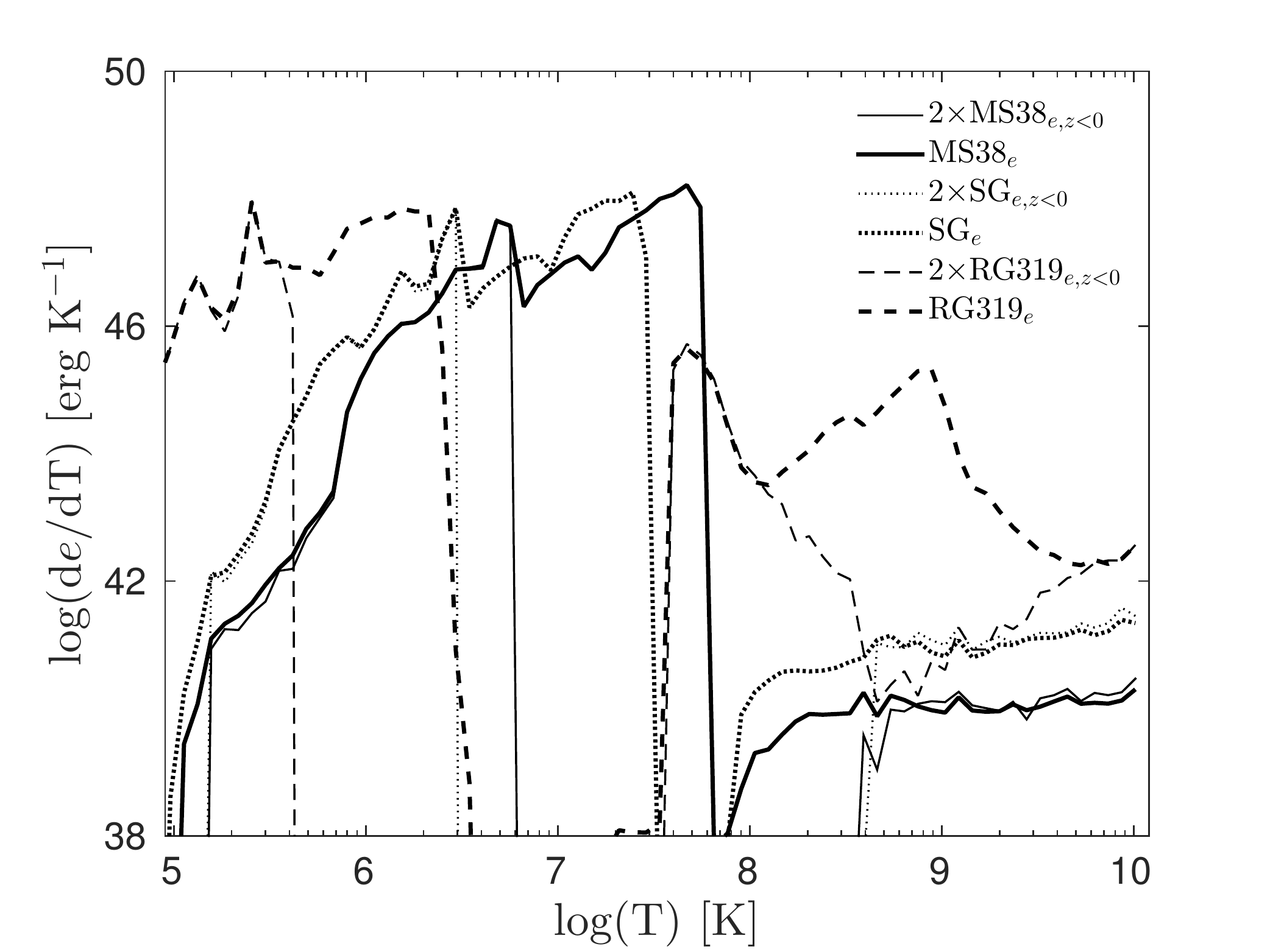}
        &
        \includegraphics[width=0.49\textwidth]{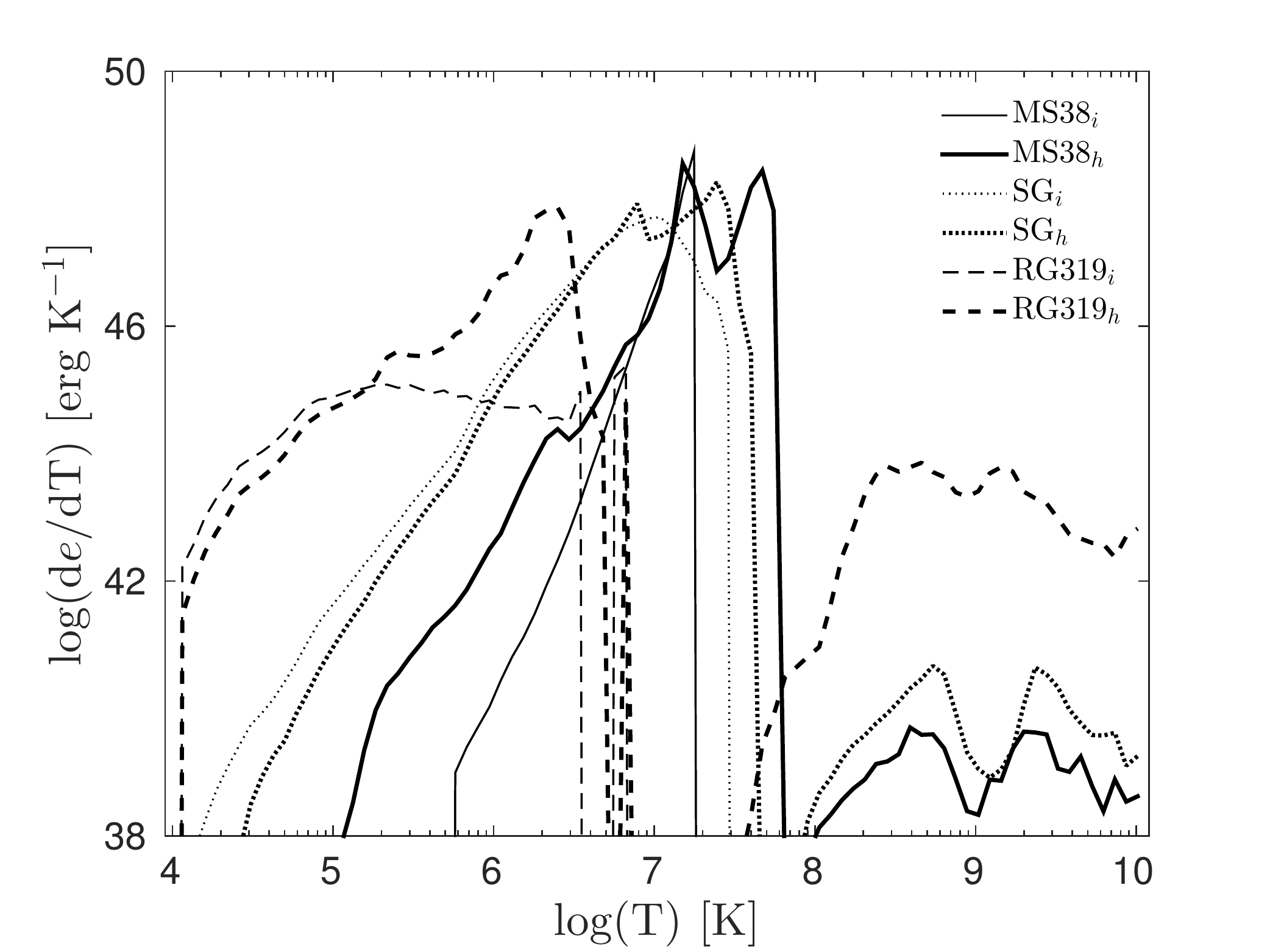}
    \end{tabular}
    \caption
    {Temperature distribution of internal energy for a subset of our
      model binary systems. For both panels, the MS model MS38 is
      shown with a solid line, the subgiant SG is shown with a dotted
      line, and the red giant RG319 is shown with a dashed line. In
      the left panel, the supernova ejecta internal energy is shown at
      the time when the increase in the ejecta's internal energy is
      near its highest point, $t_h$. Here the thin lines correspond to
      the ejecta material largely unaffected by interaction with the
      companion (twice the amount of the ejecta internal energy
      located in the lower half of the computational domain), while
      the thick lines correspond to the whole ejecta material. The
      right panel shows the bound and unbound companion material at
      the initial time, $t_i$ (thin lines), and at $t_h$ (thick
      lines), with $t_h$ defined as for the left panel. Note that the
      temperature is shown in log scale; see text for
      discussion.\label{fig:eFnTemp} }
\end{center}
\end{figure*}

Fig.\ \ref{fig:eFnTemp}
shows the internal energy distributions as a function of temperature
for our selected subset of model binary systems for the ejecta (left
panel) and the stellar envelope (right panel). The energy
distributions are shown at the simulation time, $t = t_h$, when the
increase of the internal energy due to shock heating is at its maximum
value. To estimate the amount of energy deposited by the reflected
shock (in the ejecta), we calculated the internal energy of the ejecta
material above and below the mid-plane of the computational domain,
$z=0$ cm, corresponding to the vertical position of the explosion
center. The difference between the internal energies of the ejecta
material in those two regions is equal to the energy produced by
heating due to the reflected shock. As for the heating of the stellar
envelope, the amount of internal energy deposited by the transmitted
shock can be found by simply subtracting the internal energy of the
stellar material at a given time from the initial internal energy of
the envelope.

A number of comments regarding the distributions of internal energy
shown in Fig.\ \ref{fig:eFnTemp} are in order. The effects of heating
by the reflected shock are the strongest in the case of the MS38 model
for temperatures between $T \approx 6 \times 10^{6}$ K and $T \approx
7 \times 10^{7}$ K. No internal energy is seen at the initial time in
this model (shown with a thin solid line in the left panel of
Fig.\ \ref{fig:eFnTemp}) in this range of temperatures. The reflected
shock is also capable of producing a substantial amount of
high-temperature ($T > 1\times10^8$ K) shocked ejecta, but only in the
case of low-density RG companion (shown with a thick dashed line in
the left panel of Fig.\ \ref{fig:eFnTemp}). The material located at
temperatures greater than approximately $1 \times 10^8$ K and below $3
\times 10^{8}$ K are the ejecta that swept around the companion shown
in Fig.\ \ref{fig:ms38_morph}(a). The ejecta material present at still
higher temperatures corresponds to the supernova's forward and reverse
post-shock regions (and is largely unaffected by the interaction with
the companion).

In the case of the heated companion material (shown in the right panel
in Fig.\ \ref{fig:eFnTemp}), the strongest heating occurs in the MS38
and SG models, while the RG experiences relatively weaker
heating and to lower temperatures. However, it is the RG's envelope
that experiences the relatively greatest amount of shock heating at
temperatures above $1 \times 10^{8}$ K. This hot companion material is
associated with the conical shock that develops along the center line
of the ejecta hole (cf.\ Section \ref{s:ms38results}). The properties
of this material are somewhat uncertain given the assumed symmetry in
our simulations.

Table \ref{t:residualEnergies}
\ctable[
    cap = radiation flash table,
    caption = {Residual internal energy, e$_{r,48}$\tmark, and characteristic, energy-weighted temperature of the residual internal energy,
      $T_{\mathrm{r}}$, for the supernova ejecta and companion material for a subset of our binary models.},
    label = t:residualEnergies,
    star,
    ]
    {l c c c c}
    {
      \tnote{The residual internal energy, e$_{r,48}$, is given in the units of $10^{48}$ erg.}
    }
    {\FL
      Model &            \multicolumn{2}{c}{Ejecta}               &            \multicolumn{2}{c}{Envelope}            \NN
            & $e_{\mathrm{r},48}$ [erg]  & T$_{\mathrm{r}}$ [keV] & $e_{\mathrm{r},48}$ [erg] & T$_{\mathrm{r}}$ [keV] \ML
      MS38  & 6.6 & 3.8  & 5.9 & 4.1  \NN
      SG    & 5.3 & 1.8  & 4.0 & 2.2  \NN
      RG319 & 4.0 & 0.14 & 2.6 & 0.21 \LL
    }
summarizes the characteristics of the energy deposited by the
transmitted and reflected shocks. In particular, the table provides
information about the residual, shock-deposited internal energy in the
ejecta and the envelope, $e_{r,48}$, and the internal energy-weighted,
characteristic temperature of the residual internal energy,
T$_{\mathrm{r}}$. We note that the shock heating of the companion's
envelope produces comparable amounts of internal energy to that of the
heating of the ejecta by the reflected shock. The reflected shock
produces between 10 per cent more energy than the transmitted shock in
the case of the MS38 system and 50 per cent in the case of the RG319
system. This implies that the thermalization process of the explosion
energy is more efficient for dense, compact companions. Also, the
characteristic temperatures of the energy deposited by the transmitted
shock are higher in the case of such compact companions. Consequently,
the spectra produced by those shocks will be somewhat harder than the
emission of corresponding reflected shocks. Assuming the radiating
plasma is in thermal equilibrium, the characteristic temperatures of
the shock-heated ejecta material produced in the process of the
supernova-companion interaction correspond to soft X-ray emission with
typical energies around 3.8, 1.8, and 0.14 keV in the case of MS, SG,
and RG companions, respectively.

The amount of the residual internal energy of the ejecta produced in
our models is by a factor of about 2.6--4.3 lower than the
semi-analytic estimate of the collision energy, provided by
\cite{Kasen+10}, of about $1.7 \times 10^{49}$ erg (using equation 11
in \cite{Kasen+10} for the explosion energy of our supernova
model). After taking into account the energy deposited in the
envelope, these factors decrease to about 1.4 and 4.1. Because the
contribution of the shocked envelope to the prompt emission will
likely be lower than its estimated residual internal energy, our
estimates provide upper limits for the energy available to the flash.

Kasen's estimates of the typical energies of the prompt emission are
in good agreement with our estimates of the characteristic,
energy-weighted temperatures of the shock-heated material in the case
of the RG model companion. Our MS model, however, produces shocked
plasmas with temperatures higher than those predicted by Kasen by a
factor of about 2. In passing, we note that admixture of the stripped,
shock-heated envelope material will make the emitted spectrum somewhat
harder compared to the spectrum of the emission radiated by the
shocked ejecta alone.

\cite{Marion+16} applied Kasen's model to observations of SN 2012cg,
and found the best agreement with his 6 \Msun\ model.  Because it is
very unlikely that a 6 \Msun\ MS star can be a binary companion of a
Chandrasekhar-mass white dwarf, we attempted to identify a binary
system with characteristics most closely matching the 6 solar mass
system of Kasen among our sample. The orbital distance of Kasen's 6
\Msun\ model, $a \approx 29$ \Rsun, places that system between our SG
and RG319 models. Additionally, his estimate of the characteristic
temperature of the prompt emission of 0.2 keV matches the
characteristics of our RG319 system most closely (especially if the
contribution of the shocked envelope is taken into account; the SG
emission is much harder with a characteristic temperature of about 2
keV). We conclude that the most likely companion of the SN 2012cg
system was a post-MS, possibly transiting to become a RG star.
\subsection{Numerical resolution effects}\label{ss:resolution}
The interaction between the supernova ejecta and the companion
produces flow features and involves phenomena that put the simulation
codes used to study this process to the test. On the one hand, the
simulation code must maintain a stellar companion in hydrostatic
equilibrium for several sound crossing times of the stellar
envelope. On the other hand, the code has to correctly capture the
fast-moving supernova shock, its interaction with the companion's
envelope, in particular heating of the envelope and the formation and
evolution of the reflected shock. Furthermore, the presence of shear
between the supernova ejecta and the companion's envelope induces
Kelvin-Helmholtz instability (KHI) and mixing, and contributes to the
process of stripping outer layers from the companion.

The use of an adaptive mesh refinement method, along with the
high-resolution shock capturing hydrodynamic method offered two ways
of controlling the quality of the numerical solution. First, the mesh
was refined in the regions containing substantial gradients of density
and pressure. This allows for capturing flow discontinuities such as
shocks and the KHI-unstable material interfaces that are the most
affected by numerical diffusion. Second, by changing the effective
mesh resolution in those regions we were able to control the actual
amount of numerical diffusion in the solution and thus its
quality. From the point of view of the current application, the
numerical diffusion will affect model predictions such as the amount
of stripped mass, contamination of the companion with the ejecta
material, and structure of the ejecta hole.

\begin{figure*}
\adjustThreePanels\ignorespaces
\begin{center}
    \begin{tabular}{ccc}
        \includegraphics[width=0.32\textwidth]{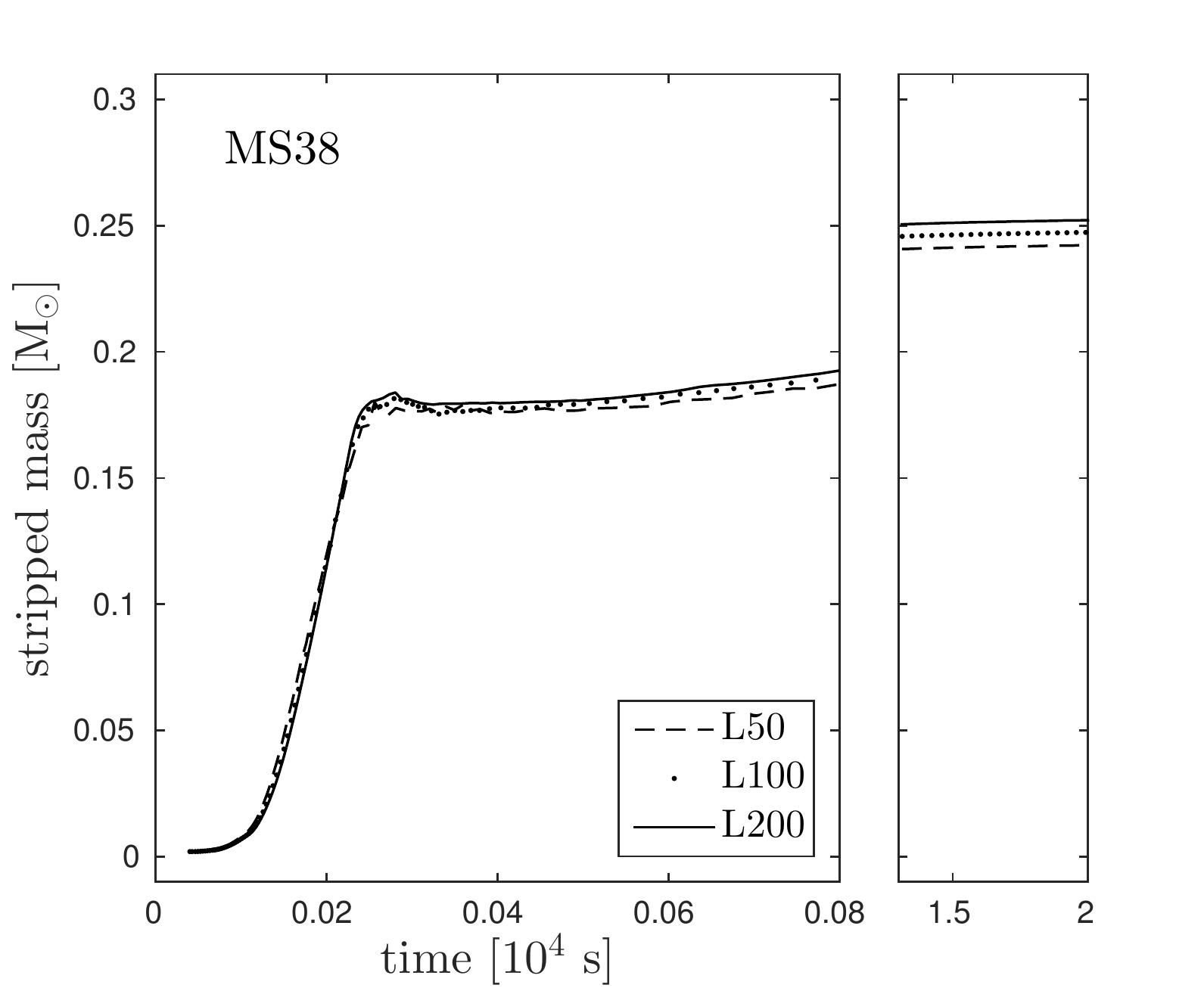}
        % \tikzImageFullDomain{figs_publish/Fig7a_w7_ms38_logd_t_19780s}{blackwhite_r}*[a][][white]*\ignorespaces
        &
        \includegraphics[width=0.32\textwidth]{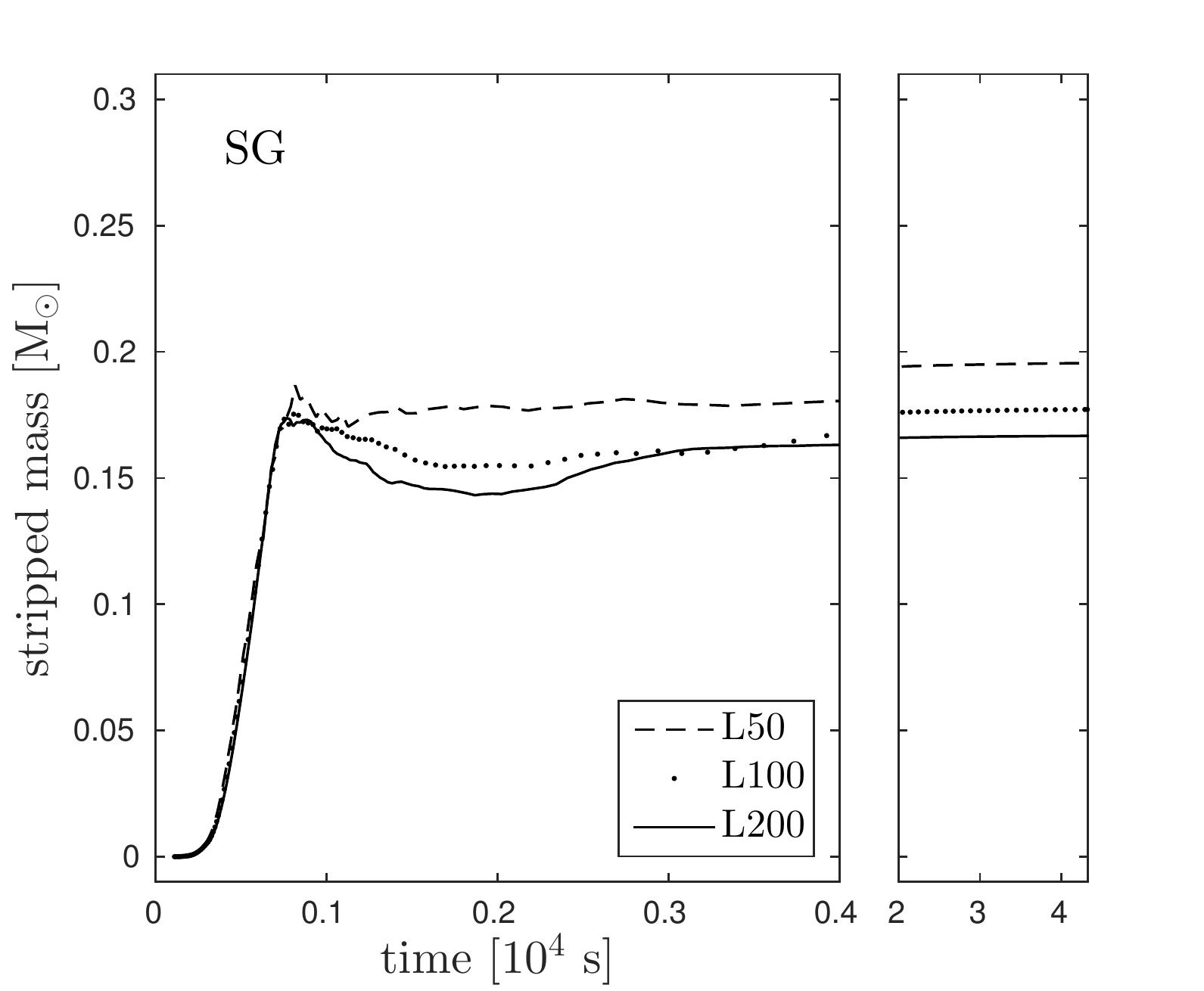}
        % \tikzImageFullDomain{figs_publish/Fig7b_w7_sg_logd_t_43200s}{blackwhite_r}*[b][][white]*\ignorespaces
        &
        \includegraphics[width=0.32\textwidth]{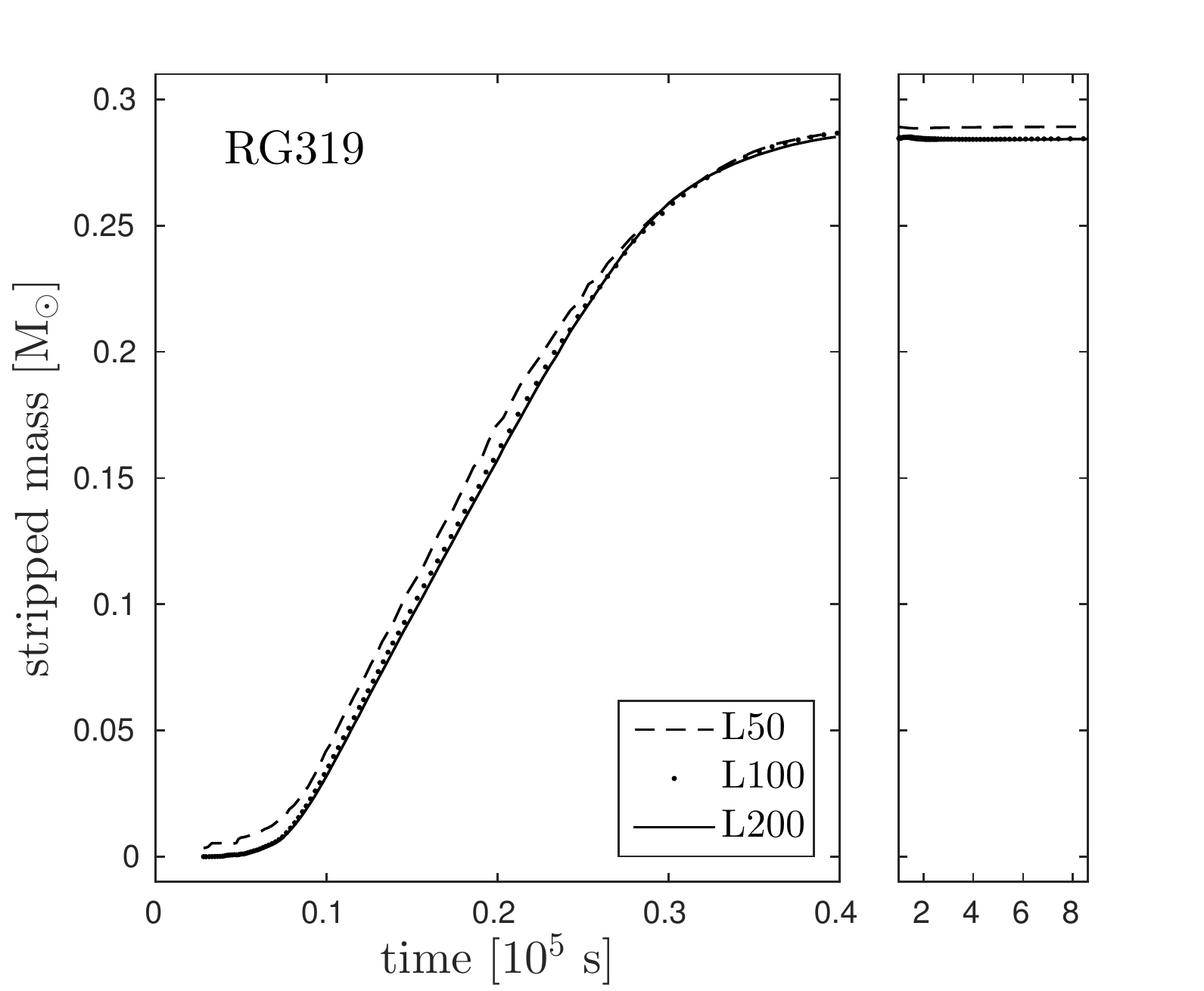}
        % \tikzImageFullDomain{figs_publish/Fig7c_w7_rg319_logd_t_842880s}{blackwhite_r}*[c][][white]*\ignorespaces
        %
    \end{tabular}
%
    % \blackLink{cb:mergerMorph0812Vert}
    % \includegraphics{figures/fig5colorbar.pdf}
% 
    \caption
    {Volume-integrated stripped companion mass as a function of time
      for a subset of model binary systems. (left panel) the main
      sequence model MS38; (middle panel) the subgiant SG; (right
      panel) the red giant RG319. Each panel is split into two
      sub-panels to show the initial behavior (left sub-panels) and
      the late-time behavior (right sub-panels). The stripped mass is
      shown with dashed, dotted, and solid lines for models with
      effective mesh resolutions of L50, L100, and L200 cells per
      companion radius, respectively. Note a relatively slow
      convergence in the amount of stripped mass in the SG
      case.\label{fig:stripped_mass_numerical}}
\end{center}
\end{figure*}

To assess the impact of numerical diffusion on the model observables
in our simulations, we performed a series of numerical experiments
varying the effective mesh
resolution. Fig.\ \ref{fig:stripped_mass_numerical}
shows the amount of stripped material as a function of time for our
basic set of binary models. For each binary type, the results are
shown in two sub-panels. In the first sub-panel we show the evolution
of stripped mass during the early, most violent, phase of the
interaction, while in the second sub panel we show the amounts of
stripped mass at late times until the final simulated time. In all
models, the amount of stripped mass increases rapidly as the ejecta
shears over the shocked envelope material, and begins to level off
around the time when the transmitted shock reaches the core of the
companion, the bound stellar material starts to expand, and mass
stripping becomes dominated by the vorticity deposited by the
supernova shock in the surface layers of the companion. The numerical
effects appear to most strongly affect the SG (middle panel in
Fig.\ \ref{fig:stripped_mass_numerical}) during that later phase as
compared to the main sequence MS38 model (left panel in
Fig.\ \ref{fig:stripped_mass_numerical}). This is most likely due to a
shallower density gradient in the envelope of the SG that allows for a
relatively greater deposition of the vorticity in its
envelope. Because the vorticity generation process involves
interaction between the gradients of density and pressure, this
process thus sensitively depends on numerical resolution.

The amount of stripped mass may decrease or increase with the
resolution, and this behavior clearly depends on the companion
type. For example, the amount of stripped mass increases with
resolution in the case of a MS star (right subpanel in the
left-hand panel of Fig.\ \ref{fig:stripped_mass_numerical}), while it
decreases for the RG model (right subpanel in the right-hand panel in
Fig.\ \ref{fig:stripped_mass_numerical}). The behavior in the case of
the SG is similar to that of the RG companion, but the
convergence is slower.

Overall, the amount of stripped mass found in the L100 and L200 models
does not differ by more than about 5 per cent, which is found in the
case of the SG model. The supernova-companion interaction problem
shares several similarities with the shock-cloud interaction
problem. The latter has been extensively studied by means of
multidimensional hydrodynamic and magnetohydrodynamic simulations. The
original work by \cite{Klein+94} considered the hydrodynamics of
shock-cloud interaction, and \cite{MacLow+94} extended the model
problem to magnetohydrodynamics. These groups found that a resolution
of at least 100 mesh cells per cloud radius was necessary in order to
correctly capture the dynamics of the shock-cloud interaction in terms
of the overall cloud shape, its integrated parameters (e.g. average
density), and the cloud destruction time. This conclusion was later
confirmed by \cite{Shin+08} in three dimensions. In the context of the
present application, the original work by \cite{Marietta+00} used a
resolution of approximately 125 zones per companion radius, and found
the amount and average dynamics of the stripped material converged at
the level of about 6 per cent. A similar accuracy of about 8 per cent
was found in the more recent three-dimensional study by
\cite{Pan+12a}, who used about 70 cells per companion radius. (A
comparable accuracy in studies using an SPH technique required at
least 500,000 particles per stellar companion \citep{Pakmor+08}.) The
estimated error of at most 5 per cent found in our L200 models is
therefore consistent with the results of convergence studies performed
by these groups.

It is important to mention that the convergence in terms of the amount
of stripped mass does not guarantee convergence in other quantities,
and likely depends on the specific characteristics of the
computational model employed. In particular, \cite{Pakmor+08}
anticipated slower convergence in terms of mixing between the ejecta
and stellar companion material due to a relatively poor performance of
SPH in resolving hydrodynamic instabilities. Also, one needs to be
particularly careful with predictions that involve relatively small
amounts of mass. The prime example here is the contamination of the
companion with the heavy elements produced by the
supernova. \cite{Pan+12a} found the maximum contamination levels of
10$^{-4}$ \Msun\ in the case of their compact helium star. This has to
be compared to the highest contamination result of 10$^{-7}$ \Msun\ in
the case of our SG. Because mesh-based, Eulerian methods are
prone to produce artificial mixing of materials, it is conceivable
that these predictions are subject to strong numerical diffusion
effects. For example, we found that the amount of contamination
decreases with resolution for our main sequence MS38 model by a factor
of 13 when the resolution is increased from L50 to L100, and by
another factor of 1.6 when the resolution is increased from L100 to
L200. In the case of our SG model, the level of contamination
decreases by a factor of about 27 when the resolution is increased
from L50 to L100, and by about 35 per cent as the resolution is
increased from L100 to L200. Clearly, more careful error estimation
studies are required when one wants to predict accurate levels of
contamination.

We found that the amounts of energy stored by both the reflected shock
in the ejecta and the transmitted shock in the envelope depend only
relatively weakly on the numerical model resolution. Specifically, the
amount of energy stored by the reflected shock in the ejecta varies by
no more than 2 per cent as the resolution increases from L100 to
L200. The convergence of the energy stored by the transmitted shock in
the companion's envelope is slowest in the case of the RG model, with
the residual energy decreasing by about 8 per cent as the resolution
is increased to 200 cells per companion radius. However, the
convergence is much faster in the case of more compact
companions. Also, we did not find a substantial dependence of the
average temperature of the residual energy on the mesh resolution in
our simulations.
\section{Summary and Conclusions}
We have studied the interaction between an exploding
Chandrasekhar-mass white dwarf with its non-degenerate companion by
means of high-resolution, hydrodynamical simulations in two
dimensions, neglecting the effects of orbital motion or spin of the
companion. We followed the evolution for a set of realistic binary
system models with MS, SG, and RG companions.

For each model binary system, we obtained a series of simulations with
progressively increasing mesh resolution in order to assess numerical
model convergence. We presented details of simulations for a
representative subset of our model binary systems. We discussed
evolution and morphology of the exploding supernova ejecta, kinematic
properties of material stripped from companion stars, and possible
contamination of companions with the the metal-rich ejecta
material. We discussed, for the first time, the amount and thermal
characteristics of the dense, X-ray emitting layers of the stellar
envelope and the ejecta heated by the shocks created in the process of
interaction of the supernova shock with the stellar envelope.

Our main findings can be summarized as follows.

\begin{enumerate}[i]

\item We found that the supernova interaction with the stellar
  companion affects the entire hemisphere of the ejecta facing the
  companion. The ejecta structure is most strongly affected in the
  part which directly impacts the companion. The visible effect of
  this interaction is the ejecta hole, a conically-shaped region
  extending behind the companion. The typical half-opening angle of
  the ejecta hole found in our simulations varied between 40 and 50
  degrees. These results are in good agreement with the findings
  reported by other groups.

\item We found the amount of mass stripped from a stellar companion to
  depend on the companion class. The envelopes of both of our RG model
  companions were completely stripped off, the finding consistent with
  the results of other groups, with both stars losing about 40 per
  cent of their initial mass. The least amount of stripped mass was
  found in the case of the SG companion, which lost only about 10 per
  cent of its initial mass.

\item The MS companion models produced mass loss in good agreement
  with the power-law relation between stripped mass and orbital
  separation of \cite{Pakmor+08}. These companions typically lost
  about a quarter of their initial mass, which is generally greater
  than the mass loss reported by other groups (our results match the
  predictions of \citealt{Pan+12a} most closely).

\item We analyzed kinematic properties of the stripped companion
  material, and found the stripped material to move with
  characteristic speeds between 500 and 700 km~s$^{-1}$. These
  characteristic speeds are uniformly lower than the values presented
  by \cite{Marietta+00}, presumably due to the lower energy of the
  supernova explosion used in our study. The numerical effects may
  also contribute to the observed discrepancy. Lower model mesh
  resolution and a different analysis method may explain much higher
  peak velocities of the stripped companion material reported by
  \cite{Pan+12a}. However, there is a good qualitative agreement
  between their velocity distributions of the stellar material and our
  results.

\item We found about $10^{-4}$ \Msun\ of hydrogen moving with speeds
  above 13\,000 km~s$^{-1}$ in our subset of model binary systems. The
  greatest amount of fast-moving hydrogen is produced in the case of
  the SG model, while the least amount is found in the case of RG
  companions. Additionally, the fastest moving hydrogen is found in
  the case of the SG companion, while the slowest hydrogen is produced
  in the process of interaction with the supernova with red giants.
  This correlation between the amount and speed of stripped hydrogen
  holds for all classes of companions also at lower velocities,
  contrary to the predictions of \cite{Marietta+00}. Therefore, and as
  originally suggested by Marietta et~al., observations of hydrogen in
  early Type Ia spectra may indeed help identification of stellar
  companion classes.

\item Our model companion stars were polluted with only small amounts
  of the ejecta material. We found the highest levels of $Ni^{56}$
  enrichment of $\approx 10^{-7}$ \Msun\ in the case of the SG
  model. Such low levels of contamination can be attributed to the
  assumed model symmetry and reduced dimensionality that prevented us
  from including the effects of orbital motion and companion's spin
  \citep[see, e.g.,][]{Pan+12a}.

\item We carefully analyzed and, for the first time, provided
  simulation-based estimates of the amounts and of the thermal
  characteristics of the shock-heated plasma expected to produce a
  flash of soft X-ray radiation during early phases of the
  supernova-companion interaction. We found the overall good
  qualitative agreement with estimates of the semi-analytic model of
  \cite{Kasen+10}. However, our models predict the energy budget
  available for the prompt emission by a factor 2--4 smaller, even
  though we also account for the energy deposited in the companion's
  envelope by the transmitted supernova shock. These numerical 
  estimates of the amounts of residual energy appear to be accurate 
  to within a few per cent in our models.

\item We estimated characteristic temperatures of the shocked ejecta
  material of 0.14 and 1.8 keV in the case of a RG and MS companion,
  respectively, in good agreement with the Kasen's model. In addition,
  we predict the SG model to produce the hardest X-ray emission of the
  shocked ejecta with average temperatures of nearly 4 keV. These
  results appear very weakly dependent on numerical model resolution
  in our simulations.

\item We also analyzed properties of the shock-heated envelope
  material of stellar companions. Although this material was not
  included in the Kasen's model, we predict that it will directly
  contribute to the prompt emission as it is transported along the
  inner surface of the ejecta hole and mixes with the shocked
  ejecta. The amount of the energy deposited in the envelope by the
  transmitted supernova shock is comparable to the energy stored by
  the reflected shock in the ejecta, while its characteristic
  temperatures are somewhat higher. The latter property implies the
  spectrum of the flash produced by both post-shock regions will be
  somewhat harder than the spectrum emitted by the shocked ejecta
  alone.

\item The plasma produced by shocks across the contact discontinuity
  will irradiate the material located across the contact surface. This
  soft X-ray emission will only be partially reflected with the
  majority of photons scattered down to lower energies \citep[see,
    e.g.,][]{Plewa95}. As a result, a softer component will be added
  to the shock-dominated prompt emission spectrum.

\item Chemical pollution of the companion star proved particularly
  sensitive to numerical diffusion effects. For example, in the case
  of our MS and SG models, the contamination of the companion star
  increased by 60 and 35 per cent, respectively, after we doubled the
  mesh resolution to 200 cells per stellar radius. This strongly
  suggests that accurate predictions of the companion's chemical
  enrichment may require simulations with 200 or more cells per
  companion's radius. The amount of stripped material appears less
  dependent on mesh resolution, and converges with accuracy better
  than about 10 per cent in models with 200 cells per stellar radius.

\end{enumerate}

Future research should focus on obtaining numerically converged,
three-dimensional models of the supernova-companion interaction to
provide reliable estimates of the stripped mass, especially in the
case of MS and SG companions. Evolved for sufficiently long times,
such models would also help to better understand the process of
pollution of companion stars with metal-rich ejecta material. Models
of the prompt emission should consider the energy budget that takes
into account the shocked ejecta as well as the shocked envelope
material, which is advected along the inner surface of the ejecta hole
and mixed with the ejecta.
\section*{Acknowledgments}

TP was partially supported by the NSF grant AST-1109113 and the DOE
grant DE-SC0008823. This research used resources of the National
Energy Research Scientific Computing Center, which is supported by the
Office of Science of the U.S. Department of Energy under Contract
No. DE-AC02-05CH11231. The software used in this work was in part
developed by the DOE Flash Center at the University of Chicago.  The
data analysis was performed in part using
MATLAB\textsuperscript{\textregistered} \citep{MATLAB+15}, and VisIt
\citep{Childs+12}. This research has made use of NASA's Astrophysics
Data System Bibliographic Services.
\label{lastpage}
\bibliographystyle{mnras}
\bibliography{references}
\end{document}